\documentclass[twocolumn]{aastex631}
\usepackage{amsmath,amssymb,epsfig,epsfig,graphicx,xspace,microtype}
\usepackage{multirow}
\usepackage{float}
\usepackage{orcidlink}

\usepackage{natbib}
\usepackage{float}

\accepted{April 3, 2026}
\submitjournal{ApJ}

\begin{document}

\newcommand{\um}{\ensuremath{\mu\mathrm{m}}\xspace}
\newcommand{\uJy}{\ensuremath{\mu\mathrm{Jy}}\xspace}
\newcommand{\degrees}{$^{\circ}$}
\newcommand{\degsq}{\ensuremath{\mathrm{deg}^2}}
\newcommand{\ssfr}{\ensuremath{\mathrm{sSFR}}\xspace}
\newcommand{\ssfrunit}{\ensuremath{\mathrm{yr}^{-1}}}

\newcommand{\oi}{[{\ion{O}{1}}]}
\newcommand{\oii}{[{\ion{O}{2}}]}
\newcommand{\oiii}{[{\ion{O}{3}}]}
\newcommand{\sii}{[{\ion{S}{2}}]}
\newcommand{\nii}{[{\ion{N}{2}}]}
\newcommand{\mgi}{[{\ion{Mg}{1}}]}
\newcommand{\mgii}{[{\ion{Mg}{2}}]}
\newcommand{\hei}{[{\ion{He}{1}}]}
\newcommand{\neiii}{[{\ion{Ne}{3}}]}

\newcommand{\OIIIHb}{[{\ion{O}{3}}]/H$\beta$}
\newcommand{\NIIHa}{[\ion{N}{2}]/H$\alpha$}
\newcommand{\SIIHa}{[\ion{S}{2}]/H$\alpha$}
\newcommand{\OIHa}{[\ion{O}{1}]/H$\alpha$}
\newcommand{\OIIIOII}{[{\ion{O}{3}}]/[\ion{O}{2}]}
\newcommand{\NIISII}{[\ion{N}{2}]/[\ion{S}{2}]}
\newcommand{\Ha}{H$\alpha$}

\def\mpch {$h^{-1}$ Mpc} 
\def\kpch {$h^{-1}$ kpc} 
\def\kms {km s$^{-1}$} 
\def\lcdm {$\Lambda$CDM } 
\newcommand{\hMpc}{\ensuremath{\,h^{-1}\, \textrm{Mpc}}}
\def\etal {et al.}
\def\kt{\tilde{k}}
\def\mpc{{\rm Mpc}}

\def\msunyr{\hbox{$M_{\odot}~{\rm yr}^{-1}$}}
\def\deg{\hbox{$^{\circ}$}}

\newcommand{\mass}{\ensuremath{\mathrm{M_*}}}
\newcommand{\mstar}{\ensuremath{\mathcal{M}_*}}
\newcommand{\msun}{\ensuremath{M_{\odot}}}
\newcommand{\logmass}{\ensuremath{\log\,(\mass/\msun)}}
\newcommand{\logmhalo}{\ensuremath{\log\,(\mass/M_{halo})}}
\newcommand{\sfr}{\ensuremath{\psi}}
\newcommand{\sfrm}{{\rm SFR}/\ensuremath{\mass}}
\newcommand{\sfrunit}{\ensuremath{\mathcal{M}_{\sun}~\textrm{yr}^{-1}}}
\newcommand{\arcsecsq}{\ensuremath{\mathrm{arcsec}^2}\xspace}

\title{A Shocked Wind Interpretation of an Odd Radio Circle}

\correspondingauthor{Alison Coil}\email{acoil@ucsd.edu}

\author[0000-0002-2583-5894]{Alison L. Coil}
\affiliation{Department of Astronomy and Astrophysics, University of California, 9500 Gilman Dr., La Jolla, CA 92093}

\author[0000-0002-1608-7564]{David S.~N. Rupke}
\affiliation{Department of Physics, Rhodes College, Memphis, TN 38112}

\author[0000-0002-2451-9160]{Serena Perrotta}
\affiliation{Department of Astronomy and Astrophysics, University of California, 9500 Gilman Dr., La Jolla, CA 92093}

\author[0009-0007-6748-4627]{Saloni Agrawal}
\affiliation{Department of Astronomy and Astrophysics, University of California, 9500 Gilman Dr., La Jolla, CA 92093}

\author[0000-0003-1785-8022]{Cassandra Lochhaas}
\affiliation{Center for Astrophysics, Harvard \& Smithsonian, 60 Garden St., Cambridge, MA 02138}
\affiliation{NASA Hubble Fellow}

\begin{abstract}

  Odd Radio Circles (ORCs) are a new class of extragalactic object, with
  large rings of faint radio continuum emission typically spanning 100s of
  kpc; their origins are unknown.  Previous optical spectroscopy of the
  central galaxy in ORC4, a classic isolated ORC, revealed spatially-extended ionized gas with
  strong \oii \ emission and line ratios consistent with LINER emission.
  We present new Keck/KCWI+KCRM integral field
  spectroscopy covering multiple strong optical emission lines to
  measure the extent, morphology, and spatially-resolved kinematics and line
  ratios of the ionized and neutral gas in the ORC4 central galaxy.
  We find that \oii \ is the
  strongest optical emission line in this massive, old galaxy, and the \oii \
  emission is detected to larger radial extent than the other optical lines.
  The gas kinematics show strong spatial asymmetries, high velocity gradients
  ($>100$ \kms), and high velocity dispersion ($\sim200$ \kms).  The emission line
  ratios are most consistent with shock models
  with shock velocities of $\sim$200-300 \kms \ and are not fit well by stellar or AGN photoionization
  models.  These findings are consistent with a model in which
   the gas in the ORC4 central galaxy is the result of shock ionization in
  and around the central galaxy, likely due to mixing and cooling of gas associated with the event that created
  the large-scale radio ring of emission that identified this source as an ORC.

\end{abstract}

% keywords:
% Circumgalactic medium (1879), Stellar feedback (1602), Galactic winds
% 33 (572), Shocks (2086)

%%%%%%%%%%%%%%%%%%%%%%
%%%%%%%%%%%%%%%%%%%%%%

\section{Introduction} \label{sec:intro}

A new class of extragalactic astronomical sources was
discovered in 2021 named Odd Radio Circles \citep[ORCs,][]{Norris21a}.  ORCs are
large rings of faint, diffuse radio continuum emission typically
spanning $\sim$1 arcminute on the sky.  These sources were initially
discovered in a large area 1 GHz radio continuum survey using the
Australian Square Kilometer Array Pathfinder (ASKAP) telescope
\citep{McConnell16}. The ASKAP array has high angular resolution and is
sensitive to low surface brightness emission.  The pilot survey
covered 270 deg$^2$ of the sky, such that it was able to detect rare,
faint, diffuse objects that had not been previously observed.
A fourth ORC (ORC4) was later discovered in archival data taken with the
Giant MeterWave Radio Telescope (GMRT) at 325 MHz,
while additional ORCs were discovered in later
ASKAP \citep{Koribalski21, Filipovic22, Gupta22} and MeerKAT
data \citep{Lochner23, Koribalski23, Koribalski24, Norris25}.

ORCs exhibit large, limb-brightened rings of radio continuum emission,
with lower surface brightness emission in the interior.  The total
number of known ORCs or ``candidate'' ORCs now numbers over ten, and
roughly half are isolated sources.  These isolated ORCs have galaxies
at their centers with red optical and infrared colors, implying dusty
or old stellar populations, with elliptical shapes and photometric
redshifts of $z\sim0.3-0.6$ \citep{Norris22}.  At these redshifts the
physical extent of the radio rings corresponds to several hundred
kiloparsecs (kpc) in size.

Multiple physical scenarios have been proposed for the origin of the
radio emission in ORCs, including Galactic supernovae remnants,
double-lobed radio galaxies seen from the side, star-forming ring
galaxies, interacting galaxies, extreme galaxy mergers, or a galaxy
virial shock \citep{Norris21a, Dolag23, Yamasaki24}. However, the most
likely scenarios involve a shock either from an outflowing galactic
wind or a blast wave driven by merging supermassive black holes
\citep{Koribalski21, Coil24}, or ORCs are remnant lobes of powerful radio
galaxies that have been reenergized by passing shocks
\citep{Shabala24}. Follow-up MeerKAT radio data on ORC1 reveals that
the large-scale radio emission is due to aged synchroton emission, and
radio polarization data show magnetic field lines that are consistent
with an expanding shell \citep{Norris22}.

Keck/KCWI observations of the one ORC discovered in the northern
hemisphere, ORC4, revealed spatially extended ionized gas observed in
\oii \ emission, as well as weak Mg II and [Ne III]
emission \citep{Coil24}.  The ionized gas extends to a radius of
$\sim$20 kpc, and there is a strong velocity gradient across
the \oii \ nebula, as well as a high velocity dispersion
($\sigma\sim200$ \kms).  SED fitting reveals that the central galaxy
is massive and old, with log $M/M_\odot = 11.3$ and a stellar age of 6 Gyr,
with a burst of star formation $\sim$1 Gyr prior. The
\oii \ equivalent width (EW) is highly elevated and
is an order of magnitude higher than in local
massive, early-type galaxies. The central galaxy has a weak AGN that
is detected in radio emission, though \citet{Coil24}
note that the \oii \ luminosity of the extended nebula
is $\sim$100 times higher than
that of typical AGN with similar radio continuum luminosities, and the
best fit SED contains only a 3\% AGN contribution at optical
wavelengths. However, they can not rule out an AGN origin for the ionized gas 
with information from \oii \ emission only.

Gemini/GMOS long-slit spectra \citep{Rupke24} reveal that 
of the first three ORCs discovered that are isolated and have central
galaxies (ORCs 1, 4, 5), all have massive central galaxies with old
stellar populations and little to no on-going star formation.
Additionally, all have LINER-like optical emission that could be
powered by shocks and/or AGN, and all three host low-luminosity,
radio-quiet AGN.  However, the ORC4 central galaxy has much stronger
line emission than the central galaxies of ORC1 or ORC5, by at least a factor
of 10 in \oii \ and \oiii \ and $\sim$10 in H$\alpha$,
within the GMOS slit aperture.

LINER emission can be due to either an AGN or shocks, though by
comparing with shock models \citet{Rupke24} concluded that the
primary ionization source in ORC4 is likely shocks due to the
mechanism that created the large-scale radio ring.  \citet{Coil24} created
numerical simulations of a three-dimensional outflowing galactic wind
driven into the circumgalactic medium and found that after the wind in
the central galaxy shuts off, the forward shock from the wind can
continue to move outward to large scales of $\sim$200 kpc on
timescales of $\sim$750 Myr.  At the same time, shocked wind gas
interior to the contact discontinuity behind the forward shock falls
back towards the galaxy to smaller scales, extending tens of
kpc.  The wind previously shocked by the reverse shock expands to fill
the under-pressurized region between the contact discontinuity and the
galaxy, creating a turbulent, energetic medium with additional shocks
as it interacts with gas in and around the galaxy, on scales of tens
of kpc, similar to the observed \oii \ emission in ORC4.  The
turbulence is expected to lead to relatively high velocity dispersion
($\sigma > 150$ \kms). \citet{Coil24} conclude that the
ionized gas from such a scenario could be observed as \oii \ emission in
the ORC4 central galaxy.

However, with blue/UV integral field spectroscopy
covering a single strong emission line, \oii,
it is not possible to draw firm conclusions about the physical origin of the
large-scale radio emission and ionized gas in ORC4.  IFU data of
longer wavelength, strong optical emission lines are necessary to
determine whether the ionized gas results from an AGN or
the presence of shocks.
In this paper we present new Keck/KCWI+KCRM data to spatially map
emission in ORC4 at longer optical wavelengths to test the wind shock
theory for the origin of this ORC.  We measure the spatial extent and
spatially-resolved kinematics and line ratios of the ionized gas as
traced by multiple emission lines in and around the central galaxy in
ORC4.  With KCWI+KCRM we are able to detect extended, low surface
brightness emission that is unobservable with GMOS long-slit data and
spatially resolve a suite of strong optical lines in ORC4 including
\oii, H$\beta$, \oiii, [OI], H$\alpha$, \nii, and \sii.  Only with
integral field spectroscopy covering this suite of lines can we
distinguish {\it nuclear} LINER activity from spatially-extended shocks, to
determine the origin of the ionized gas in the ORC4 central galaxy.

%%%%%%%%%%%%%%%%%%%%%%%%%%%%%%%%%%%%%%%%%%%%%%%%% 

\begin{figure*}[ht!]
\centering
\includegraphics[width=5.6in]{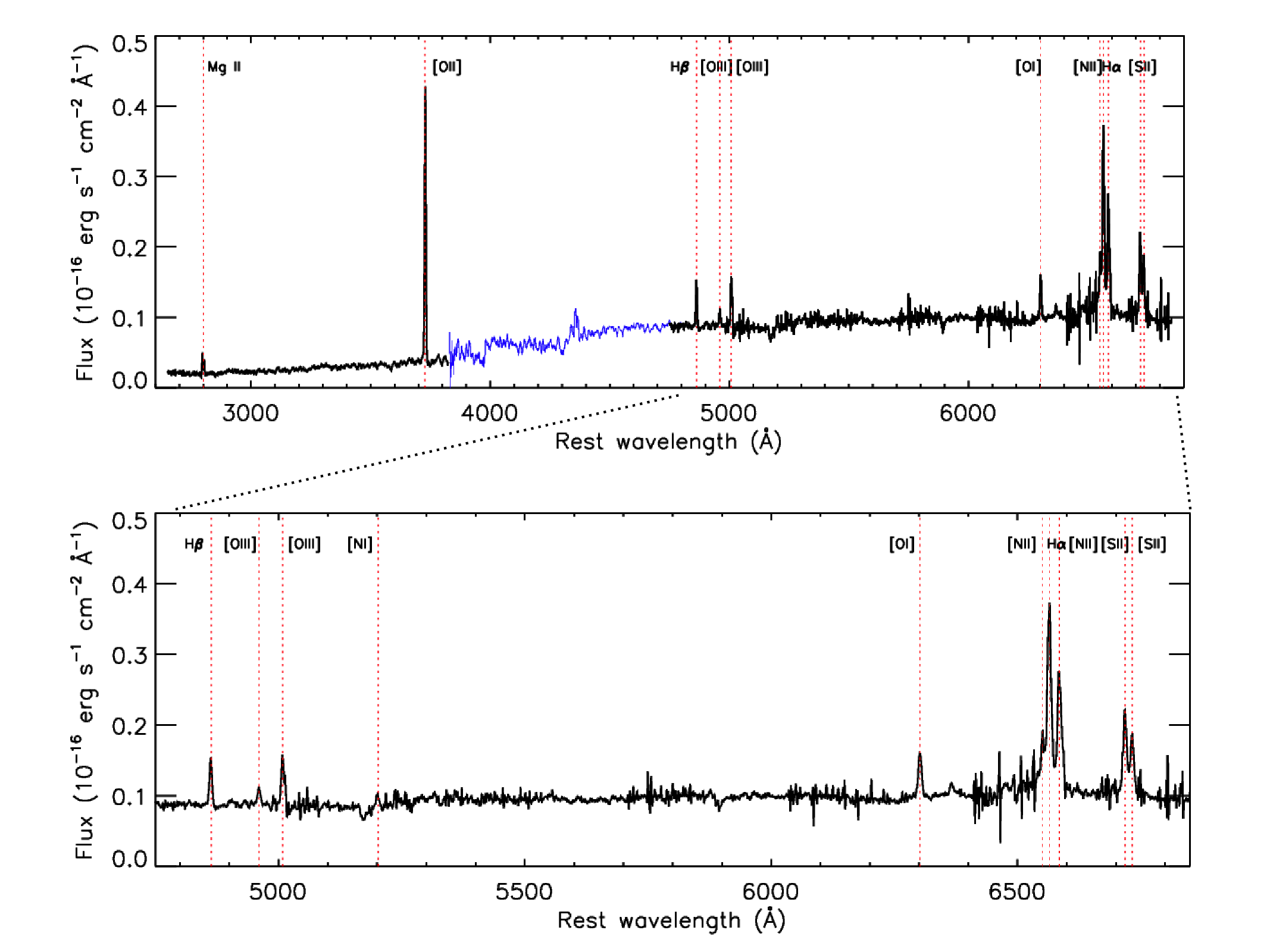}
\caption{Spatially-integrated optical spectrum of the ORC4 central
  galaxy, showing stellar and line emission integrated in a circular
  aperture of 2$\arcsec$ diameter centered on the galaxy.  The upper
  panel shows the spectra from both the blue and red arms of
  KCWI+KCRM (in black), while the lower panel shows the KCRM spectrum in
  greater detail.  The gap in wavelength coverage from $\sim$3900-4700 \AA \
  is intentional (see text for details); the blue line shows the scaled GMOS
  long-slit spectrum of this galaxy from \citet{Rupke24} in the wavelength range
  not covered by KCWI+KCRM.
  Emission lines are marked with red dotted vertical lines and labelled.
  \oii \ is the strongest line in the optical spectrum.
}
\label{fig:spectrum}
\end{figure*}

%%%%%%%%%%%%%%%%%%%%%%%%%%%%%%%%%%%%%%%%%%%%%%%%% 

%%%%%%%%%%%%%%%%%%%%%%
%%%%%%%%%%%%%%%%%%%%%%

\section{Observations, Data, and Spectral Fitting} \label{sec:obs-red}

\subsection{Observations}

We observed ORC4 with the Keck Cosmic Web
Imager \citep[KCWI,][]{Morrissey18}, which has been upgraded to include
the Keck Cosmic Reionization Mapper \citep[KCRM,][]{McGurk24} red arm,
on the Keck II telescope on July 1, 2024.  The upgrade expanded the
wavelength coverage of KCWI to include 5600 \AA\ to 10800 \AA, such
that the entire optical spectrum can be observed.  We used the
medium slicer with a central blue wavelenth of 4600 \AA\ and a central
red wavelength of 8400 \AA; while this results in a gap in the
wavelength coverage between the blue and red arms it allows us to
observe from Mg II at 2800 \AA\ to \sii \ at 6731 \AA\ simultaneously.
The bandpass of the red arm does not allow for continuous coverage from the
dichroic at 5600 \AA\ to 10000 \AA, and we prioritized detection of emission from
H$\beta$ to \sii \ at red wavelengths such that light just redward of
the dichroic is missed, as it does not fall on the detector, given the width
of the bandpass.

We used the BL and BR gratings and 2$\times$2 spatial binning, which results
in a spaxel size of 0.29$\arcsec$ $\times$0.69$\arcsec$ and a field of
view of 16$\arcsec$ $\times$ 20$\arcsec$ per pointing. This configuration
has a spectral resolution of $R=$1800, corresponding to a velocity resolution
of 167 \kms \ at the dichroic. We used
exposures of 22 minutes on the blue side and 4$\times$5 minutes on the red
side and dithered half a slice (0.35\arcsec) between exposures.  We
exposed for a total of 3.67 hours at a position angle of 100 degrees,
near the parallactic angle, at an airmass of 1.0 to 1.2. The weather
was clear and the seeing was $\sim$0.9$\arcsec$.

\subsection{Data Reduction}

We reduced the data using the Python version of the KCWI Data
Reduction Pipeline (DRP, v1.2.0), the KSkyWizard Python package
(https://github.com/zhuyunz/KSkyWizard), and the IFSRED IDL library
\citep{Rupke14}.  Initial sky subtraction was performed using the DRP
with manual selection of a sky mask region within each pointing.
KSkyWizard was used to improve the sky subtraction, using principal
component analysis (PCA) techniques, and the telluric correction.
The standard star BD+26d2606 was used for flux calibration.
Wavelengths were converted to vacuum.

Following these pipeline stages, we resampled the data onto
0.29$\arcsec$ $\times$ 0.29$\arcsec$ spaxel grids using the routine
IFSR\_KCWIRESAMPLE and rotated the mosaiced data cube to a position
angle of 0 degrees. The resulting data cube has dimensions of 65
$\times$ 70 spaxels, covering 19$\arcsec$ $\times$ 20$\arcsec$.
Assuming a $\Lambda$CDM Planck 2018 cosmology \citep{Planck20} 
 with H$_0$ = 67 \kms \ Mpc$^{-1}$, $\Omega_m$ = 0.315, and
$\Omega_{\Lambda}$ = 0.685, the
physical dimensions of a single KCWI pointing correspond to 113 kpc
$\times$ 122 kpc at the redshift of the source, $z$=0.4515.

% 5.96 kpc per arcsec
% each spaxel is 1.74 kpc

%%%%%%%%%%%%%%%%%%%%%%%%%%%%%%%%%%%%%%%%%%%%%%%%% 

\begin{figure*}[ht!]
\centering
\includegraphics[width=5.6in]{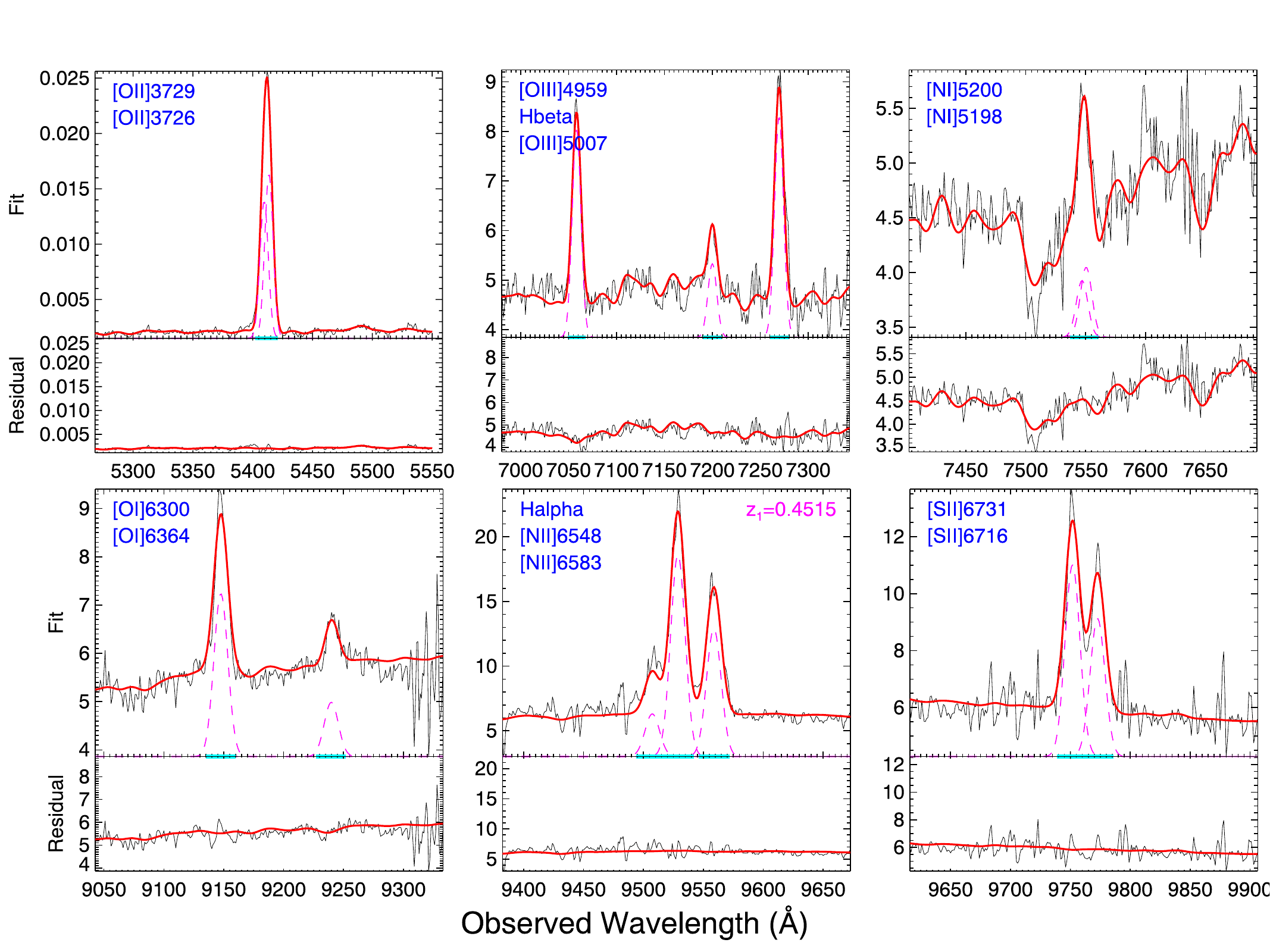}
\caption{Combined stellar continuum and emission line fits (in red, upper panels) showing
  \oii, \oiii, H$\beta$, [\textrm{N}~\textsc{i}], [\textrm{O}~\textsc{i}], 
 \nii \,
 H$\alpha$, and \sii \ in the central spaxel of the ORC4 central galaxy. The data are shown in black.  The lower panels show the stellar continuum fit after subtracting the line emission, shown with dotted pink lines, and cyan regions indicate the wavelengths used to identify and fit the emission lines.
}
\label{fig:emfits}
\end{figure*}

%%%%%%%%%%%%%%%%%%%%%%%%%%%%%%%%%%%%%%%%%%%%%%%%% 

\subsection{Spectral Fitting}

We modelled the spectrum in each spaxel using the IFSFIT library
\citep{Rupke14} in IDL. It incorporates PPXF \citep{Cappellari12} to
fit the stellar continuum, and MPFIT \citep{Markwardt09} to fit
Gaussian profiles to the emission lines. IFSFIT first masks emission
line regions and fits the stellar continuum, then simultaneously fits
the emission lines in the continuum-subtracted spectrum.
The continuum spectrum is fit with the 
BPASS \citep{Stanway18, Byrne22} high-resolution stellar
population synthesis model assuming solar or twice solar metallicity,
and additive 7th-order Legendre
polynomials to account for residuals from imperfect calibration such
as scattered-light or residual sky background.
In fitting the emission lines we fit the
Mg II 2800\AA \ doublet, \oii \ 3726\AA, \oii \ 3729\AA, H$\beta$,
\oiii \ 4959\AA,
\oiii \ 5007\AA, [\textrm{N}~\textsc{i}] 5198\AA, [\textrm{N}~\textsc{i}] 5200\AA, \hei \ 5876\AA, [OI] 6300\AA, \nii \ 6548\AA, H$\alpha$, \nii \ 6583\AA,
\sii \ 6716\AA, and \sii \ 6731\AA \ lines,  allowing a maximum of
one Gaussian component for each emission line.  We fit
the KCWI blue detector and KCRM red detector data separately; on the blue side 
the kinematics of the \mgii \ doublet is tied to that of \oii, while on the red
side the kinematics of each line is tied to that of H$\alpha$. Therefore the
kinematics of \oii, for example, may differ from that of H$\alpha$, while 
the kinematics of \oiii \ and H$\alpha$ will agree. 
 Model line profiles were
convolved with the spectral resolution before fitting.  The \oii \
doublet is unresolved in our observation, and the \oii \ 3729/3726 \AA \ flux
ratio is fixed to 1.2, corresponding to an electron density of 400
$cm^{-3}$ \citep{Pradhan06}.  Varying this ratio from a value of 0.8 to 1.4 results
in a systematic offset in the center of the line fit which corresponds to a velocity
offset of 49 \kms. Following the spectral fitting,
emission lines with a significance of
less than 3$\sigma$ in their flux are set to zero. From these resulting fits we
created flux and kinematics maps for each emission line. To measure
stellar kinematics, we additionally performed a continuum-only fit to
the continuum data cube. Emission lines were masked and the continuum
was fitted with the same stellar population model and Legendre polynomials
described above, and the stellar population models were convolved with the spectral resolution.

%%%%%%%%%%%%%%%%%%%%%%%%%%%%%%%%%%%%%%%%%%%%%%%%%

\begin{deluxetable}{lrrr}
  \tablecaption{Line Luminosities and Equivalent Widths\label{tab:lums}}
%  \tablewidth{1.0\columnwidth}

  \tablehead{\colhead{Line} & \colhead{$L_{obs}$} & \colhead{$L_{dust-cor}$} & \colhead{Restframe} \\
    \colhead{} & \colhead{(10$^{41}$ erg s$^{-1}$)} & \colhead{(10$^{41}$ erg s$^{-1}$)}& \colhead{EW (\AA)}}
  \colnumbers

  \startdata
      {\mgii} 2800  &    0.47$\pm0.09$  & 4.1$\pm0.8$  &  21$\pm5$ \\
      {\oii} 3727   &    6.3$\pm0.06$   & 35$\pm3.5$   &  58$\pm2$  \\ 
      H$\beta$      &    1.2$\pm0.09$   & 4.3$\pm0.3$  &  5.4$\pm0.4$ \\
      {\oiii} 4959  &    0.33$\pm0.01$  & 1.2$\pm0.02$  &  1.5$\pm0.04$ \\
      {\oiii} 5007  &    0.98$\pm0.07$ & 3.4$\pm0.2$  &  4.6$\pm0.3$ \\ 
      {\textrm{N}~\textsc{i}} 5200   &    0.32$\pm0.1$ & 1.0$\pm0.2$  &  1.5$\pm0.3$ \\
      {\oi} 6300   &     1.0$\pm0.06$  & 2.7$\pm0.2$  &  4.2$\pm0.3$ \\
      {\nii} 6548   &    1.1$\pm0.02$  & 2.7$\pm0.1$ &  4.1$\pm0.1$ \\
      H$\alpha$     &    5.1$\pm0.14$   & 12.8$\pm0.3$ &  19$\pm0.6$ \\  
      {\nii} 6583   &    3.2$\pm0.14$   & 7.9$\pm0.4$  &  12$\pm0.6$ \\
      {\sii} 6716   &    2.1$\pm0.07$  & 5.0$\pm0.2$  &  8.1$\pm0.3$ \\
      {\sii} 6731   &    1.4$\pm0.06$  & 3.4$\pm0.2$  &  5.6$\pm0.3$ \\
  \enddata

  \tablecomments{The \mgii \ and \oii \ luminosities listed are for the unresolved doublets. Line luminosities are given for the observed flux and the flux corrected for extinction within the galaxy; see text for details. }
  
\end{deluxetable}

%%%%%%%%%%%%%%%%%%%%%%%%%%%%%%%%%%%%%%%%%%%%%%%%% 

%%%%%%%%%%%%%%%%%%%%%%
%%%%%%%%%%%%%%%%%%%%%%

\section{Results} \label{sec:results}

\subsection{Stellar Continuum Fit}

Figure \ref{fig:spectrum} shows spatially-integrated spectra spanning
the full optical wavelength range, integrated over the central 2$\arcsec$ diameter of the ORC4 central galaxy where the S/N is highest.
\oii \ is clearly the strongest emission line in the ORC4 central galaxy
spectrum, and a high ratio of \nii/H$\alpha$ is evident.

The systemic redshift of $z = 0.4515$ is determined by the fit to the
stellar continuum, which can be seen in Figure \ref{fig:spectrum}. 
Although the discontinuous wavelength coverage prevents 
detection of the 4000\AA\ break, the long-slit GMOS spectrum of ORC4
presented in \citet{Rupke24}, and shown in blue in Figure \ref{fig:spectrum}
within the wavelength region not covered by KCWI, 
clearly reveals an old stellar population for which \citet{Rupke14} measure
D$_n$(4000) = 1.30. SED modeling of the ORC4 central galaxy in
\citet{Coil24} and \citet{Rupke24} also indicates an old, massive
stellar population with \logmass = 11.3, which had a burst of
star formation $\sim$1 Gyr ago. From the KCWI data presented here we
measure a velocity dispersion for the stars in the spatially-integrated
spectrum of $\sigma = 177 \pm22$ \kms. No significant rotation is observed in the
stellar velocities across the ORC4 central galaxy.

\subsection{Emission Line Luminosities and Equivalent Widths}

Fits to emission lines in a spatially-integrated spectrum covering the central
4$\arcsec$ (24 kpc) diameter of the ORC4 central galaxy are shown in Figure
\ref{fig:emfits}.  Total luminosities are given in Table 1, which lists both the
luminosities derived from the observed flux and those corrected for dust extinction within the
ORC4 central galaxy (see below). The luminosities are calculated by summing over the
emission line fits in spaxels with S/N$>3$ for each line, and errors are estimated from fits to the spatially-integrated spectrum.  The systematic uncertainty due to the dust correction is larger than the statistical errors listed in the table. 
Table 1 also lists the restframe equivalent widths (EW).

We estimate the color excess
$E(B-V)_{gas}$ to be $0.39 \pm0.03$ from the Balmer decrement, assuming case B,
the \citet{Cardelli89} extinction law, and $R_V = 4.05$. To measure the Balmer
decrement we sum the flux in H$\alpha$ and H$\beta$ over all spaxels with
$S/N>3$ in each line. The error on the color excess is estimated by propogating the statistical error on the Balmer decrement from the fits to the spatially-summed H$\alpha$ and H$\beta$ line fluxes. We do not perform spatially-resolved dust corrections based on the Balmer decrement measured in each spaxel, as the H$\beta$ flux is not high when measured for individual spaxels and as such this ratio is uncertain. The spaxel-to-spaxel variation in the the color excess, using those spaxels in the central 10 kpc diameter where the S/N is highest, is 0.17, which is substantially greater than the error of 0.03 derived from the spatially-summed fluxed.

As can be seen from the table, the luminosity of the \oii \ doublet
is higher than that of any other lines, including H$\alpha$,
and the restframe EW of \oii \ is very high, 58 \AA, while the EW of
H$\alpha$ is 19 \AA\ and the EW of \nii \ 6583 \AA\ is 12 \AA.
As noted by \citet{Coil24}, the \oii \ EW in the ORC4 central galaxy is an
order of magnitude higher than what is typically found in red, early-type
galaxies \citep{Yan06}.  The \sii \ 6716 \AA / \sii \ 6731 \AA \ ratio is found to be 1.43, corresponding to the low electron density regime of $n<10 \ cm^{-3}$.

%%%%%%%%%%%%%%%%%%%%%%%%%%%%%%%%%%%%%%%%%%%%%%%%% 

\begin{figure}[t!]
\centering
\includegraphics[width=3.5in]{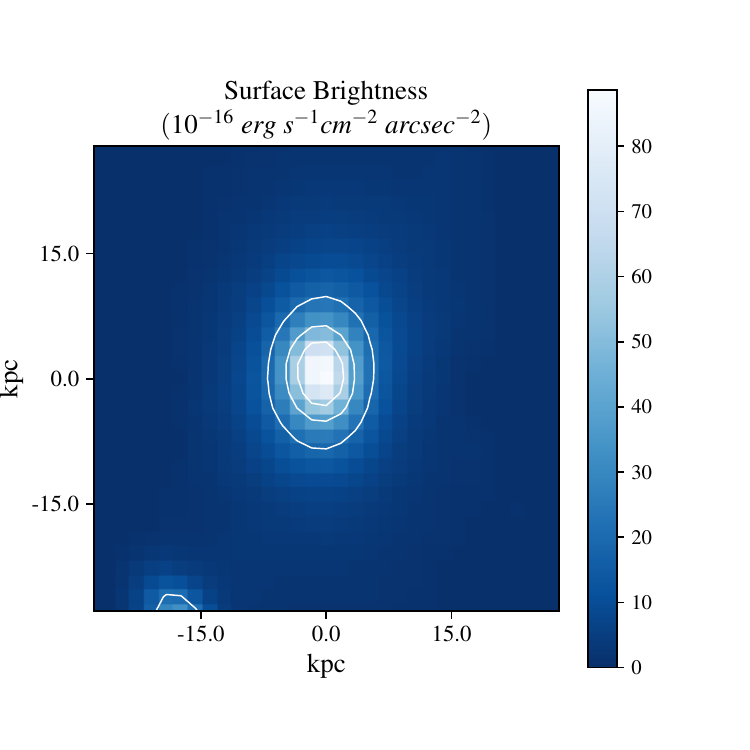}
\caption{Surface brightness map of the stellar continuum emission in the ORC4 central galaxy, integrated over the observed wavelength range 7600\AA $-$ 9000\AA.
  Contours show
    surface brightness levels of 20, 40, and 60  $\times$ 10$^{-16} \
  erg \ s^{-1} \ cm^{-2} \ arcsec^{-2}$.
}
\label{fig:sb_cont}
\end{figure}

\begin{figure*}[ht!]
\centering
\includegraphics[width=5.6in]{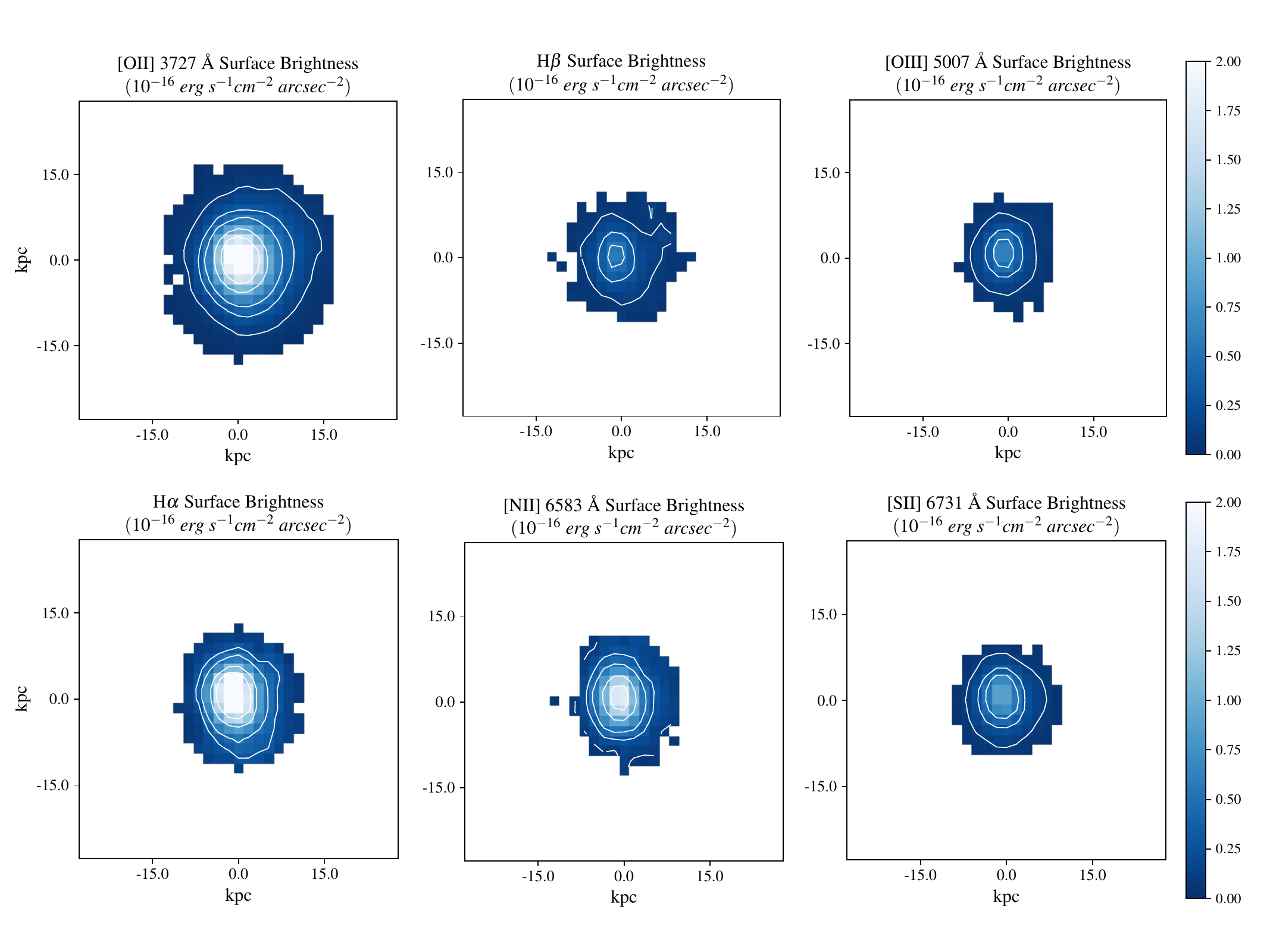}
\caption{Surface brightness maps of six of the strongest emission lines in the
  ORC4 central galaxy; spaxels with $S/N>3$ are shown. The upper left panel shows emission from the \oii \ doublet, while
  the rest of the panels show emission from single lines. Contours show
  surface brightness levels of 0.1, 0.3, 0.5, 1.0, and 1.5 $\times$ 10$^{-16}
  \ erg s^{-1} cm^{-2} arcsec^{-2}$.
}
\label{fig:sb_maps}
\end{figure*}

%%%%%%%%%%%%%%%%%%%%%%%%%%%%%%%%%%%%%%%%%%%%%%%%% 

%%%%%%%%%%%%%%%%%%%%%%%%%%%%%%%%%%%%%%%%%%%%%%%%% 

\begin{figure}[ht!]
\centering
\includegraphics[width=3.5in,trim={0.5in 0 0 0},clip]{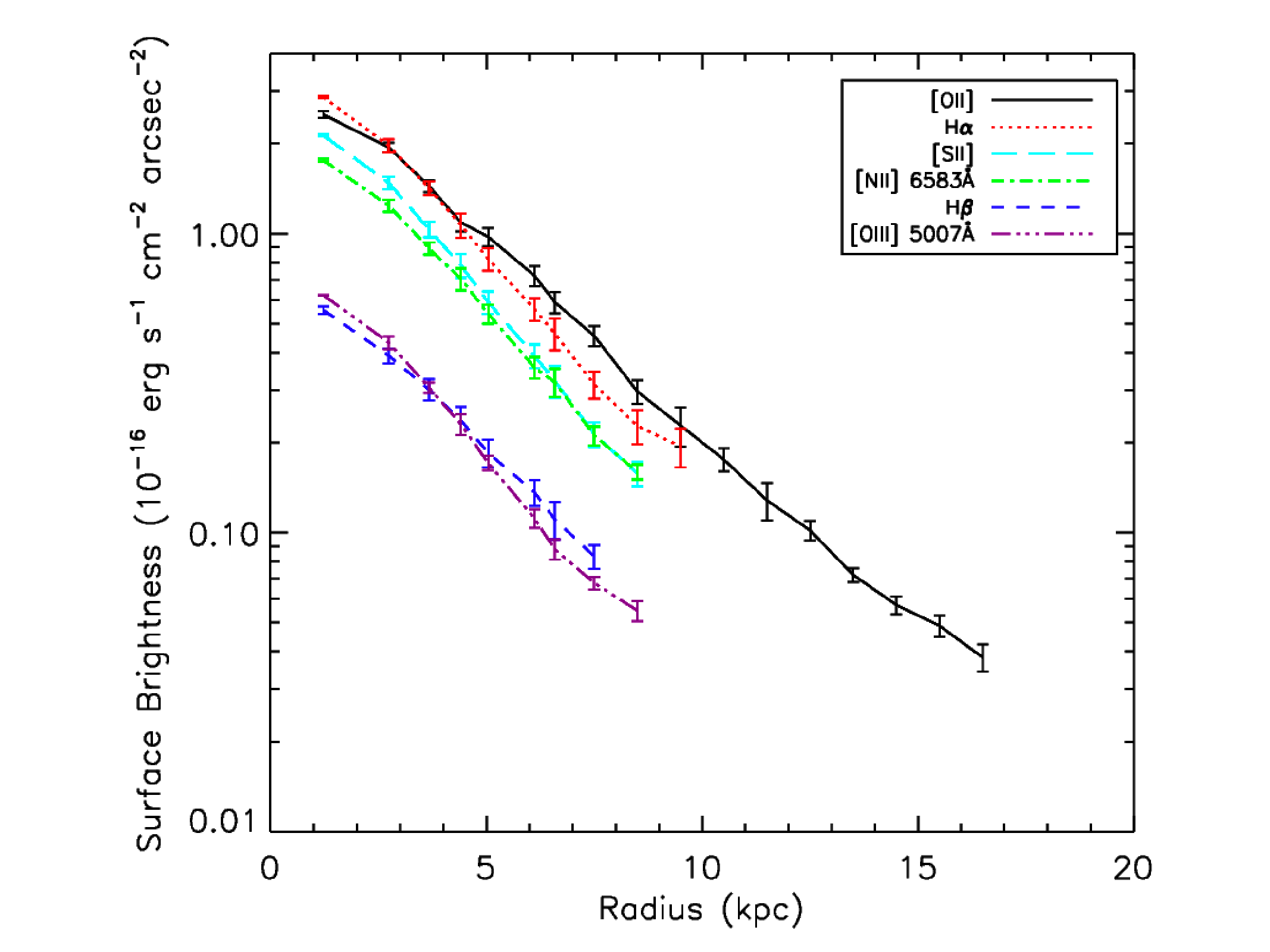}
\caption{Surface brightness radial profiles are shown for the \oii \ doublet,
  H$\alpha$, \sii \ doublet, \nii \ 6583\AA, H$\beta$, and \oiii \ 5007\AA \ emission lines, as a function of distance from the center of the ORC4 central galaxy.
}
\label{fig:rad_profile}
\end{figure}

%%%%%%%%%%%%%%%%%%%%%%%%%%%%%%%%%%%%%%%%%%%%%%%%% 

%%%%%%%%%%%%%%%%%%%%%%%%%%%%%%%%%%%%%%%%%%%%%%%%% 

\begin{figure*}[ht!]
\centering
\includegraphics[width=6.7in,trim={0 3in 0 3in},clip]{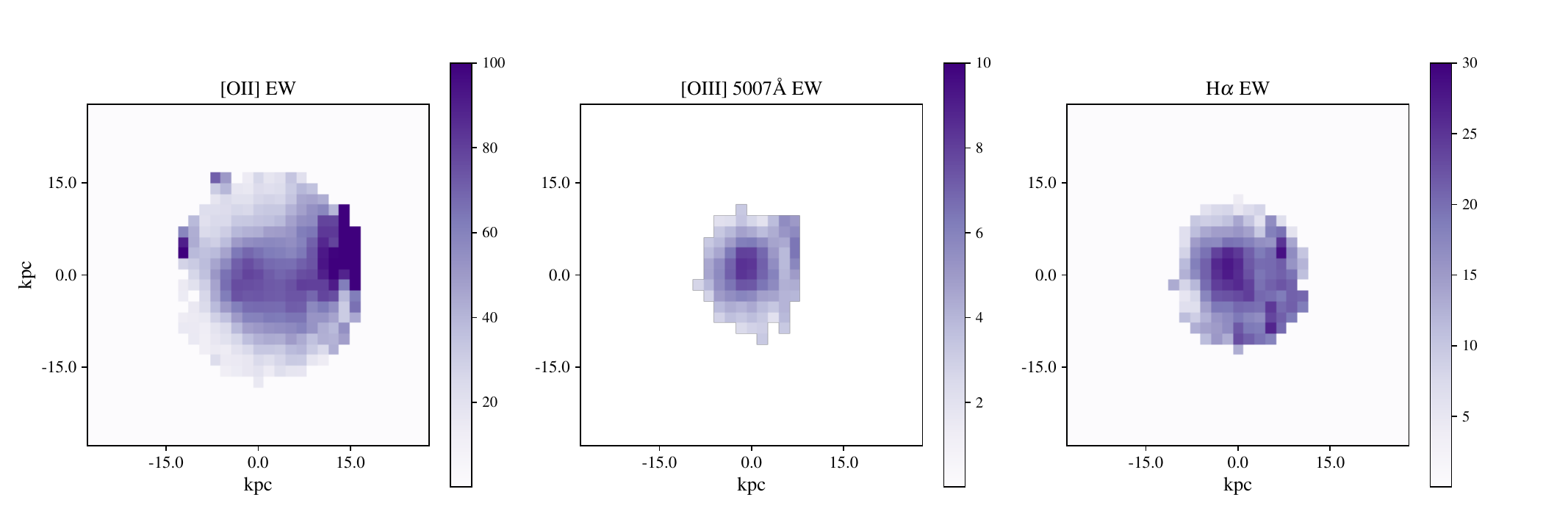}
\caption{Restframe equivalenth width (EW) spatial maps of three of the strongest
  optical emission lines in the
  ORC4 central galaxy; the left panel shows the EW for the \oii \ doublet.
  The color bar range is unique for each line; \oii \
  has the highest restframe EW (in units of \AA), followed by H$\alpha$. 
}
\label{fig:ew_maps}
\end{figure*}

%%%%%%%%%%%%%%%%%%%%%%%%%%%%%%%%%%%%%%%%%%%%%%%%% 

\subsection{Surface Brightness Maps and Radial Profiles}

Figures \ref{fig:sb_cont} and \ref{fig:sb_maps} show surface brightness
maps of the stellar continuum and several strong emission lines
in the ORC4 central galaxy. Only spaxels with $S/N>3$ are displayed.
North is up and east to the left
in all maps, which have been rotated to a position angle of 0.
The stellar emission is elongated along the north-south direction. The
edge of the source seen in the lower left region of Figure  \ref{fig:sb_cont} is a nearby galaxy; it is
36 kpc away projected on the sky and is at the same redshift as ORC4
\citep{Coil24}. 
The optical spectrum of this galaxy reveals an old stellar population with
D$_n$(4000) = 2.34, no detected emission lines, and a stellar mass 0.8 dex
less than the central galaxy in ORC4 \citep{Rupke24}.

The surface brightness maps of line emission in the ORC4 central
galaxy (Figure \ref{fig:sb_maps}) reveal that the \oii \ emission  is detected to larger radius
than the other strong emission lines.  This is further illustrated in Figure \ref{fig:rad_profile},
which shows the detected radial surface brightness profiles of the same ions at in Figure \ref{fig:sb_maps}. The profiles are created by summing the surface brightness
in radial bins for each emission line, including only spaxels with
a $S/N>3$. The profiles for both \oii \ and \sii \ show the summed
doublet in each case, while the other profiles are for individual emission
lines. Beyond a radius of 5 kpc the \oii \ surface brightness is higher at
a given radius than the surface brightness of the other emission lines;
the next brightest lines are H$\alpha$ and the \sii \ doublet. While H$\alpha$ is
relatively bright it is not detected beyond a radius of 9 kpc due to contamination from a sky
line, which makes the fits unreliable at lower $S/N$. 
While most lines are
detected out to radii of $\sim$7-10 kpc, \oii \ is observed extending to a radius of
17 kpc. We are able to detect the \oii \ line to lower surface brightness than the other emission lines as it is not contaminated by sky lines. As discussed in \citet{Coil24}, this large physical extent is very
unusual for \oii \ emission in massive, early-type galaxies.
While the other emission lines show roughly spherically symmetric emission
in the ORC4 central galaxy,
the \oii \ emission in Figure \ref{fig:sb_maps} extends at lower surface
brightness levels towards the northwest, seen most clearly in the lowest contour level.

Restframe EW maps of \oii, 
\oiii \ 5007\AA, and H$\alpha$ are shown in Figure \ref{fig:ew_maps}.
While not shown here, the EW maps of
the other emission lines are similar to those seen for \oiii \ 5007\AA\ and
H$\alpha$. The maximum restframe EW of the \oiii \ 5007\AA\ emission is 8.4 \AA,
seen near the center of the galaxy.  The restframe H$\alpha$ EW maps peaks at
28 \AA, approximately 5 kpc east of the galaxy center. Although a few spaxels along the northwest and southwest edges also show high EW values, these are likely noise as the elevated measurements in these regions are confined to only a handful of spaxels.

The restframe EW map for \oii \ is more complex.  The EW in the center
of the source peaks 3.5 kpc east with a value of 79 \AA; the
EW is slightly lower (72 \AA) at the very center of the galaxy, similar
to what is seen for \oiii \ 5007 \AA\ and H$\alpha$. Additionally, the
highest EW is seen along the
west/northwest edge of the \oii \ nebula, where the EW reaches values of
$>100$ \AA.  This elevated EW appears to be real as
there are more than 20 spaxels with a substantially higher EW in this region
spanning $>$10 kpc.
The contours in the surface brightness map for \oii \ in the upper left
panel of Figure \ref{fig:sb_maps} show this extension in the \oii \ emission
to the northwest at low surface brightness.  The \oii \ restframe EW along
the outer region
of the rest of the nebula is $\sim$20 \AA\ at a radius of 15 kpc. 

The increase in restframe EW along the west/northwest edge was not apparent
in the original KCWI observations presented by \citet{Coil24}, which were
substantially shallower. The original dataset had a total exposure time of
20 minutes, compared to 3.7 hours of integration in the present observations.
While the deeper data do not reveal \oii \ emission at larger radii, they do
uncover new features in the \oii \ emission that were not detected in the
original data.

 As shown in \citet{Coil24}, the \oii \ EW and radial extent
  measured in the ORC4 central galaxy are highly unusual when compared
  to similarly massive, early-type galaxies.  \citet{Pandya17}
  observed a sample of 74 early type galaxies with \logmass $>$ 11.5
  and found a median \oii \ radial extent of 2.3 kpc, with \oii \ EWs
  typically $<$ 10 \AA.  The \oii \ emission in ORC4 is therefore an
  order of magnitude higher in both EW and radial extent than in
  comparable massive ellipticals.

%%%%%%%%%%%%%%%%%%%%%%%%%%%%%%%%%%%%%%%%%%%%%%%%% 

\begin{figure*}[ht!]
\centering
\includegraphics[width=6.7in,trim={0 3.5in 0 2.9in},clip]{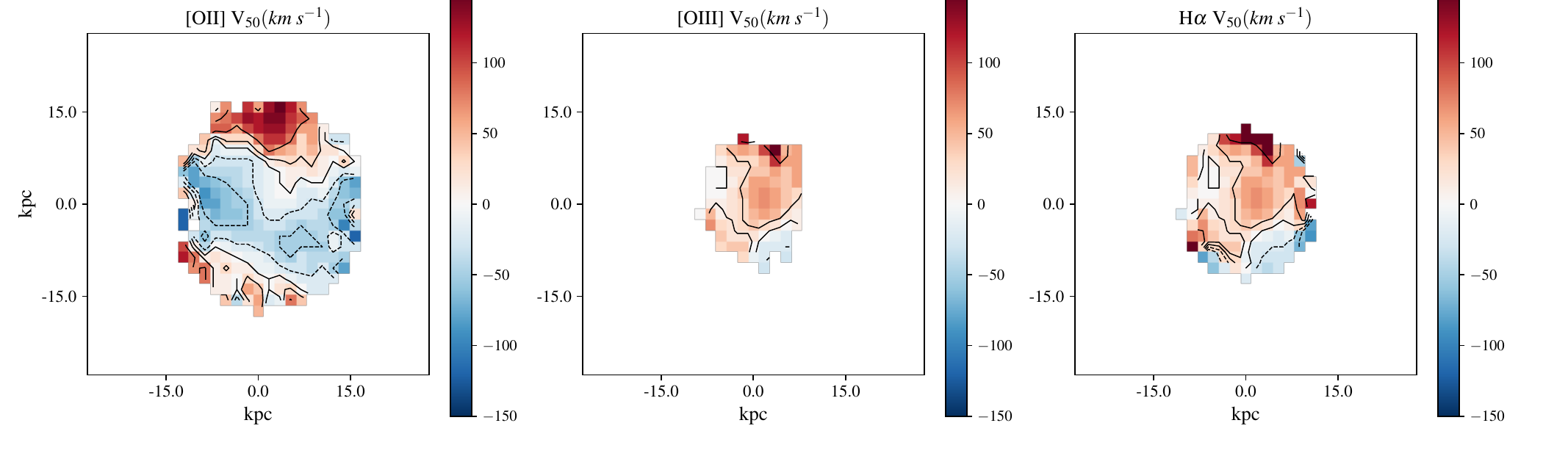}
\caption{The central velocity, $V_{50}$, measured in each spaxel for
  the \oii \ doublet (left), \oiii \ 5007 \AA \ (middle), and H$\alpha$ (right) emission lines. As discussed in the text, the kinematics of fits to the \oii \ line are not tied to those
  of the redder optical emission lines, while the kinematics of the \oiii \ and H$\alpha$ lines are tied by the emission line fitting procedure and thus agree by design.
}
\label{fig:vel_cen}
\end{figure*}

\begin{figure*}[ht!]
\centering
\includegraphics[width=6.1in,trim={0.1in 2.9in 0 2.9in},clip]{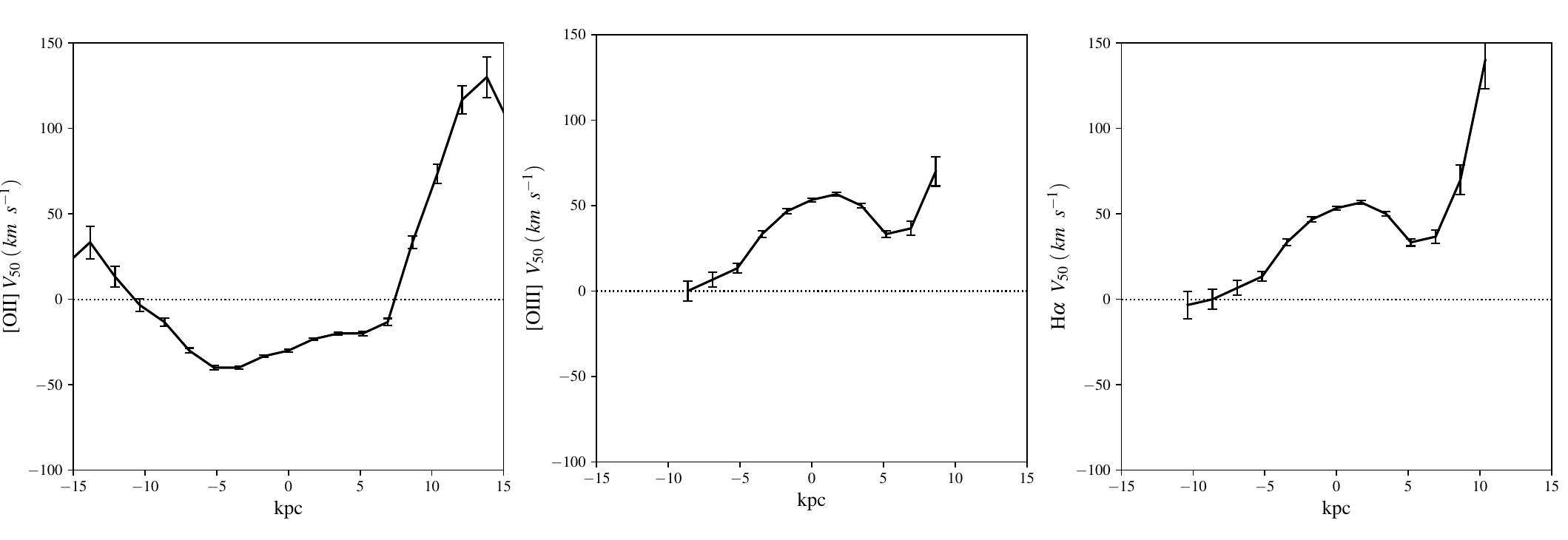}
\caption{The central velocity, $V_{50}$, measured along a north-south cut through the center of each map shown in Figure \ref{fig:vel_cen}, averaged across the central three columns, for 
 the  \oii \ doublet (left), \oiii \ 5007 \AA \ (middle), and H$\alpha$ (right) emission lines. North is towards the right in this figure.
}
\label{fig:pv}
\end{figure*}

%%%%%%%%%%%%%%%%%%%%%%%%%%%%%%%%%%%%%%%%%%%%%%%%% 

\subsection{Gas Kinematics}

We next investigate the kinematics of the extended gas in the ORC4 central galaxy.
As described above, the \oii \ line emission was fit separately from the redder 
emission lines, due to the \oii \ nebula being  detected to larger spatial extent, while
the other strong lines were fit simultaneously. This allows for detection of
differences in the kinematics of the \oii \ emission relative to the other strong
lines, if present.  In Figure
\ref{fig:vel_cen} we present the central velocity, $V_{50}$, measured in each
spaxel for the \oii \ doublet (left), \oiii \ 5007 \AA \ line (center), and
H$\alpha$ line (right).
Figure \ref{fig:pv} presents the central velocity as a function of distance
from the center of the galaxy along the north-south axis, averaging over
the central three columns in the maps shown in Figure \ref{fig:vel_cen}.
The \oii \ nebula exhibits kinematics distinct from the other emission lines.
The velocity zeropoint is set by the redshift of
the stellar continuum fit, such that these figures show the gas velocity relative
to the stars.  In the center of the galaxy the \oii \ emission is blueshifted
by $\sim$30 \kms \ relative to the stars, while the other lines show a redshift of $\sim$50 \kms.
The \oii \ emission also shows a strong, asymmetric velocity gradient
across the nebula, with highly redshifted emission in the north, $>100$ \kms,
and a maximum blueshift along a roughly east-west direction through the center
of the galaxy, with mildly redshifted gas ($\sim$25 \kms) in the south.
The other emission
lines are predominantly redshifted with respect to the stars in the ORC4 central
galaxy and show the highest redshift in the north. The H$\alpha$ kinematics show
a strongly rising redshift along the northern edge, with velocities redshifted by
more than 100 \kms. 
While the \oii \ emitting gas appears to be kinematically distinct from
the gas observed in other emission lines, the velocity difference of $\sim$80 \kms \ observed
between the central velocity measured for \oii \ relative to the other strong lines in the center of the galaxy 
is not statistically significant given that there is a systematic uncertainty of 49 \kms \
due to the assumed \oii \ doublet flux ratio, and given that the velocity resolution
of the data is 167 \kms.   There is no detectable second kinematic component in either the \oii \ or other strong line emission. It is notable that a pronounced
redshift of greater than 100 \kms \ is observed along the northern ridge in all of the strong lines.

\begin{figure*}[ht!]
\centering
\includegraphics[width=6.5in,trim={0 2.9in 0 2.9in},clip]{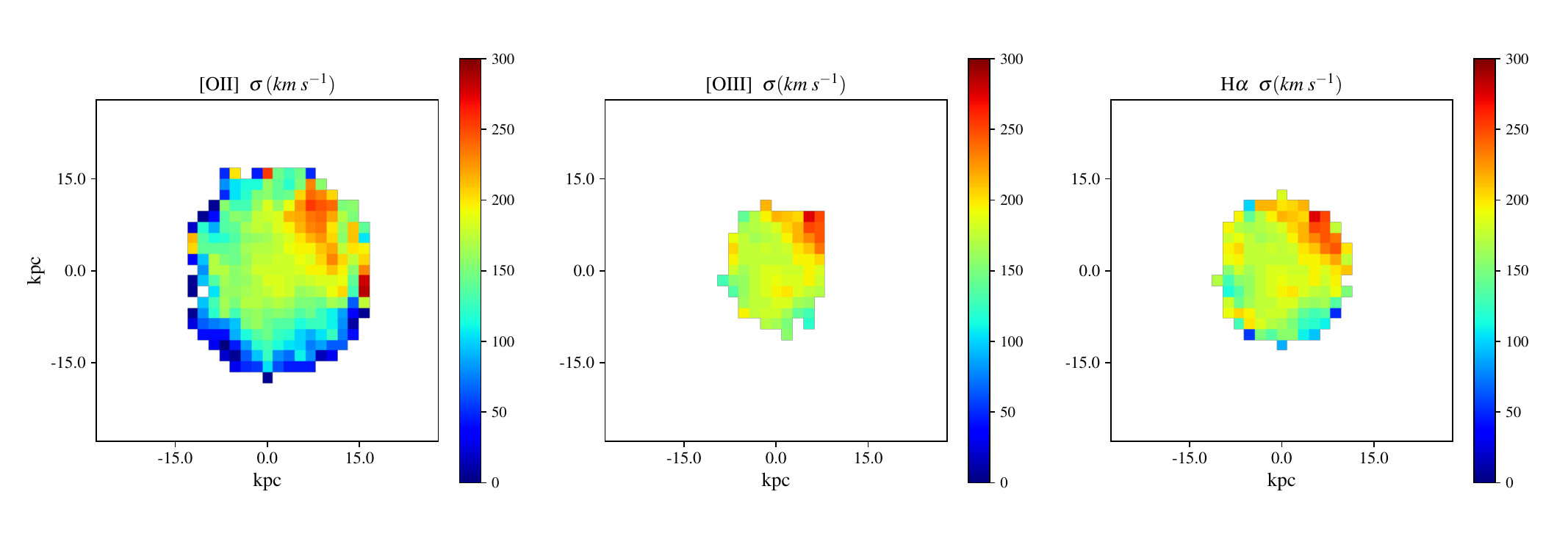}
\caption{The velocity dispersion, $\sigma$, measured in each spaxel for
  the \oii \ doublet (left), \oiii \ 5007 \AA \ (middle), and H$\alpha$ (right) emission lines. The color bar range is identical in all three panels, spanning 0 to 300 \kms. While the kinematics of the \oii \ doublet are not tied to the kinematics of \oiii \ or H$\alpha$, the spatial $\sigma$ maps look very similar.
}
\label{fig:sigma}
\end{figure*}

Figure \ref{fig:sigma} shows the velocity dispersion, $\sigma$, in each spaxel
for the gas observed in \oii \ (left), \oiii (center), and H$\alpha$ (right).
The velocity dispersion shows less variation across the different emission lines
than is observed in the central velocity maps; in particular the spatial variation in $\sigma$ in the \oii \ and the redder emission lines agrees well. Both \oii \ and H$\alpha$
show high velocity dispersion
at the center of the galaxy with $\sigma\sim180$ \kms, increasing to the northwest
to $\sim$250 \kms along the edge of the \oiii \ and H$\alpha$ nebulae.  The \oii \
nebula shows an increase in $\sigma$ in the same direction to a maximum value of 240 \kms, and a low velocity dispersion around the edges in the other directions.
The \oii \ velocity dispersion map resembles that presented in \citet{Coil24},
though the deeper data result in a noticeable less noisy map.

The median of the ratio of $\sigma / |V_{50}|$ for the \oiii \ emission is 6, with
values near 3 in the center of the galaxy, increasing to 20 near the edges.
The median $\sigma / V_{50}$ for
H$\alpha$ is 5. The median of this ratio for the \oii \ emission is 4 and
shows strong spatial variation, with elevated values ($\sim20$) towards the center of
the nebula.

The gas kinematics clearly show that the gas in the ORC4 central galaxy is not
undergoing simple rotation, and the high $\sigma$ values in particular reflect
that the gas is disturbed.  Regions of high $\sigma$ could result from
multiple distinct velocity components, though at the spectral resolution of this data
that is not observed, or it could be due to turbulence. 
This is discussed further in Section 4 below.

%%%%%%%%%%%%%%%%%%%%%%%%%%%%%%%%%%%%%%%%%%%%%%%%% 

%%%%%%%%%%%%%%%%%%%%%%%%%%%%%%%%%%%%%%%%%%%%%%%%% 

\begin{figure*}[ht!]
\centering
\includegraphics[width=6.5in,trim={0 1.5in 0 0.4in},clip]{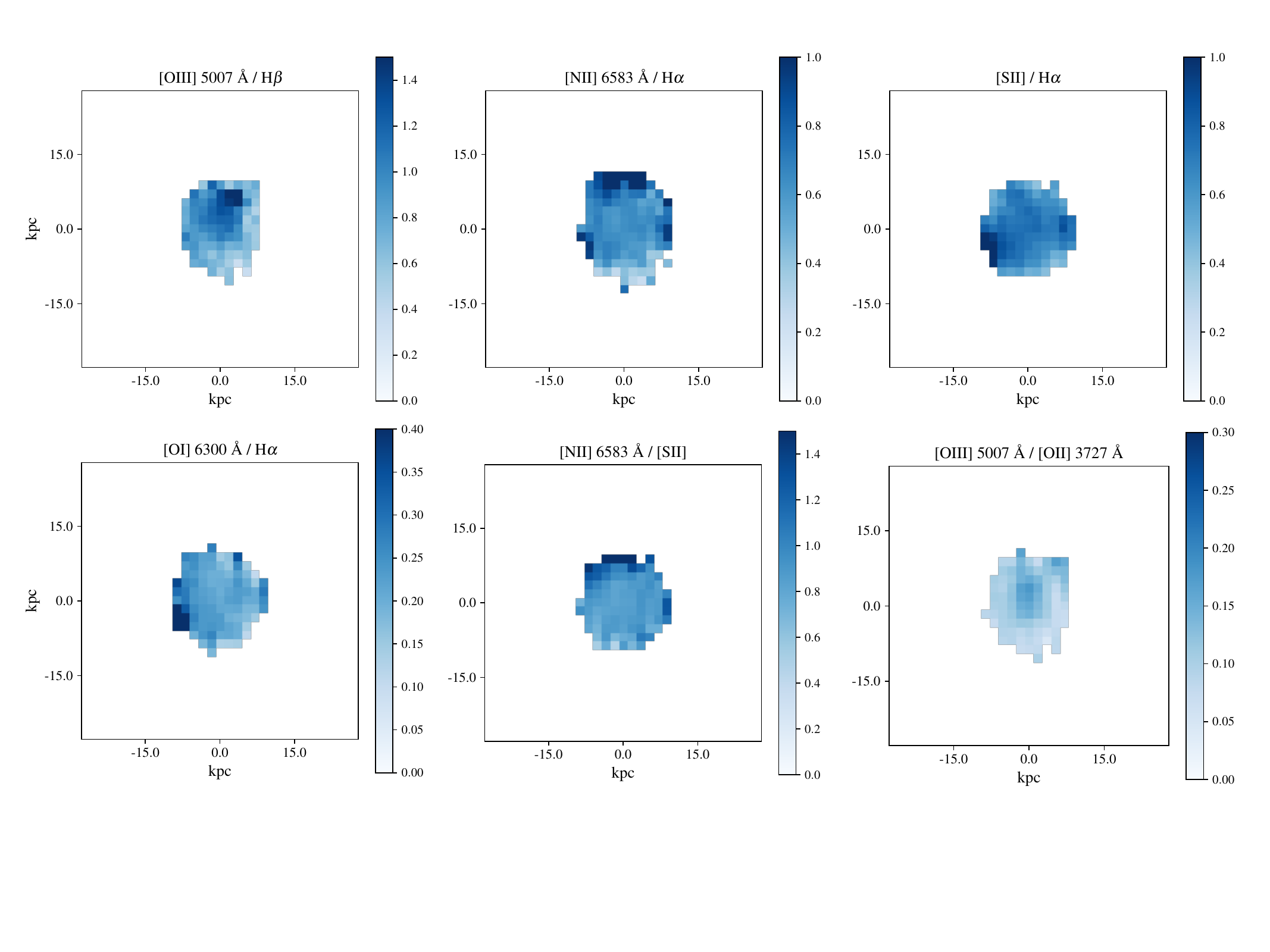}
\caption{Spatial maps of six common strong line ratios: \oiii \ 5007 \AA / H$\beta$ (upper left), \nii \ 6583 \AA / H$\alpha$ (upper center), \sii \ 6716 + 6731 \AA / H$\alpha$ (upper right), \oi \ 6300 \AA / H$\alpha$ (lower left), \nii \ 6583 \AA / \sii \ 6716 + 6731 \AA (lower center), and \oiii \ 5007 \AA / \oii \ 3727 \AA \ doublet (lower right), corrected for dust extinction in the ORC4 central galaxy. The color bar varies in each panel, to highlight the observed spatial variation within each line ratio.
}
\label{fig:line_ratio_maps}
\end{figure*}

%%%%%%%%%%%%%%%%%%%%%%%%%%%%%%%%%%%%%%%%%%%%%%%%% 

\begin{figure*}[ht!]
\centering
\includegraphics[width=6.5in]{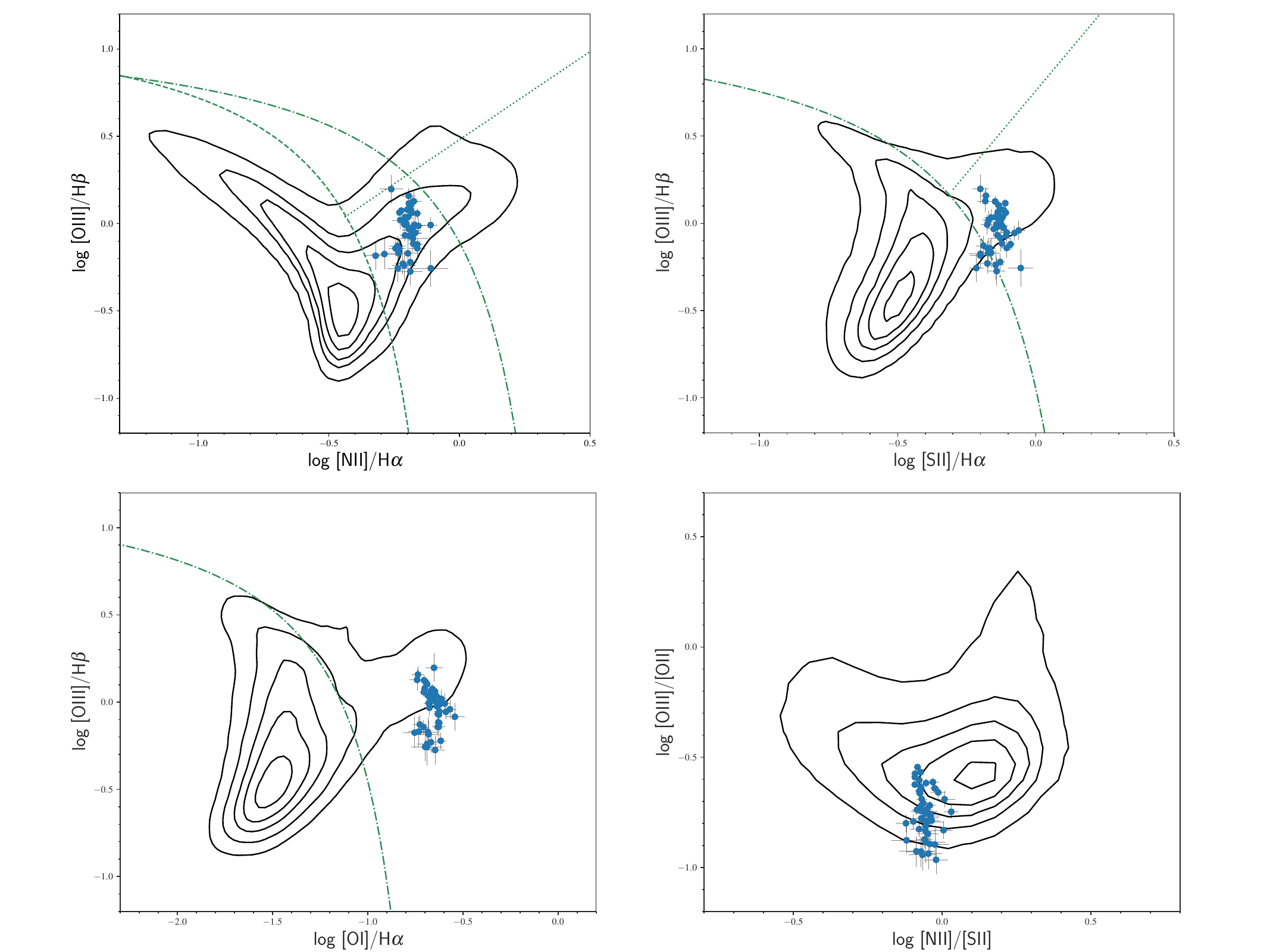}
\caption{Strong emission line flux ratio diagrams: \OIIIHb \ versus \NIIHa \
  (upper left), \OIIIHb \ versus \SIIHa \ (upper right), \OIIIHb \ versus
  \OIHa \ (lower left), and \OIIIOII \ versus \NIISII \ (lower right).
  The blue points with error bars show spaxels in the 
  ORC4 central galaxy, not corrected for dust extinction.  The green lines show
  commonly-used demarcations for starburst versus AGN versus LINER regions of
  these diagrams (see text for details). 
  Contours show SDSS galaxies for comparison; density
  contours show 10\%, 30\%, 50\%, 70\%, and 90\% of the SDSS DR7 population,
  for sources with S/N$>3$ in each of the relevant emission lines. }
\label{fig:bpt_sdss}
\end{figure*}

%%%%%%%%%%%%%%%%%%%%%%%%%%%%%%%%%%%%%%%%%%%%%%%%% 

\begin{figure*}[ht!]
\centering
\includegraphics[width=6.5in]{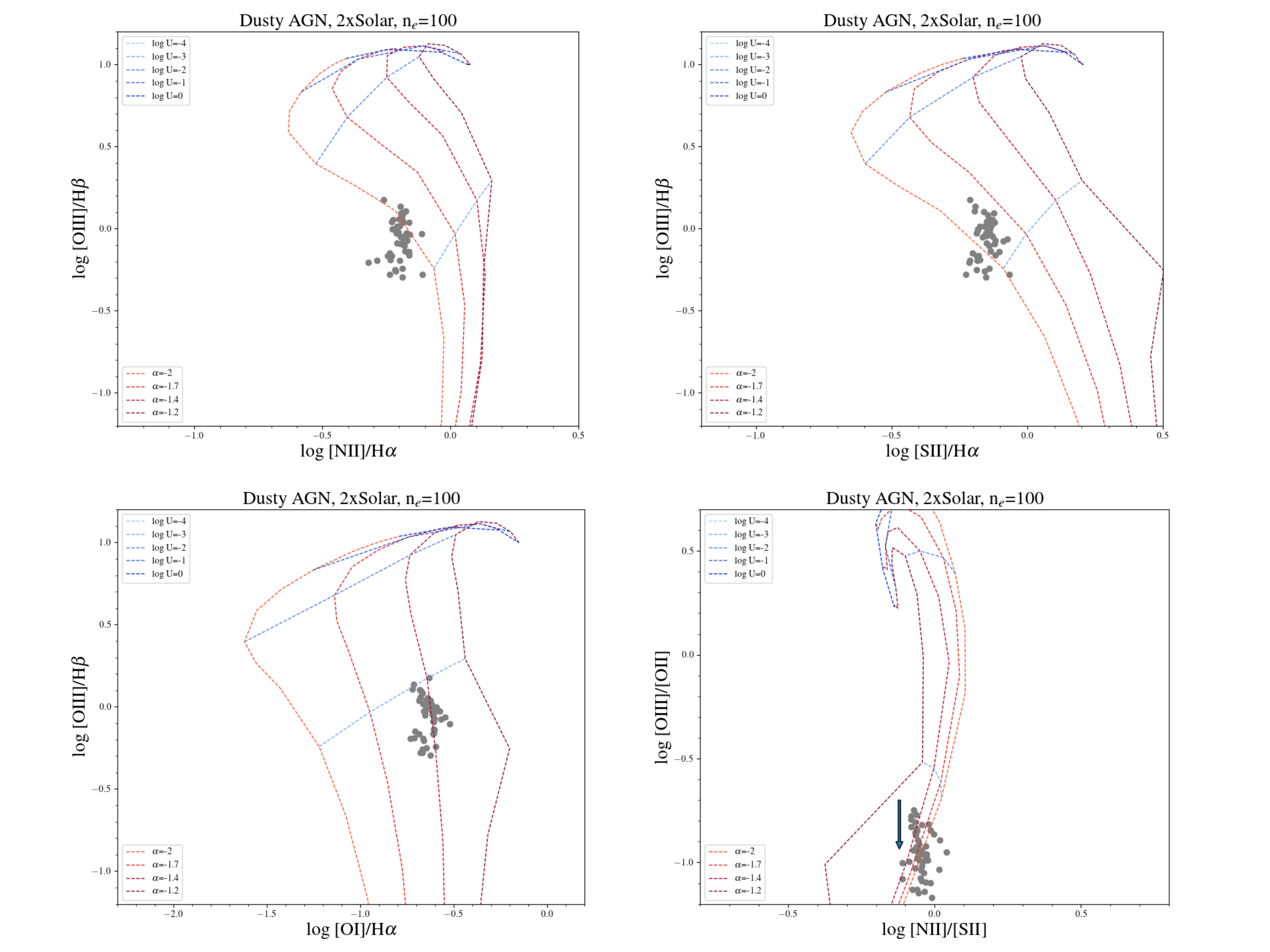}
\caption{A comparison of the extinction-corrected line ratio diagrams from
  Figure \ref{fig:bpt_sdss}, where grey points show spaxels for the ORC4 central
  galaxy, with dusty AGN model grids from \citet{Groves04} for twice solar
  metallicy and an electon density $n_e$ = 100 $cm^{-3}$. The grids show varying values of
  the AGN power-law index ($\alpha$) and ionization parameter (log $U$). 
  Here the ORC4 central
  galaxy points have been corrected for dust extinction derived from the estimated
  color excess, $E(B-V)$, in the central galaxy. The blue arrow in the lower right
  panel indicates the magnitude and direction of this correction in this panel; the
  correction is negligible for the other three panels.}
\label{fig:agn_models2}
\end{figure*}

%%%%%%%%%%%%%%%%%%%%%%%%%%%%%%%%%%%%%%%%%%%%%%%%% 

\subsection{Line Ratios} \label{sec:line-ratios}

We next investigate strong emission line flux ratios in the ORC4
central galaxy in order to better understand the ionizing source and nature
of the gas. Figure \ref{fig:line_ratio_maps} presents spatial maps of six
common strong line flux ratios:
\oiii \ 5007 \AA / H$\beta$, \nii \ 6583 \AA / H$\alpha$, \sii \ 6716 + 6731 \AA / H$\alpha$,
\oi \ 6300 \AA / H$\alpha$, \nii \ 6583 \AA / \sii \ 6716 + 6731 \AA, and
\oiii \ 5007 \AA / \oii \ 3727 \AA \ doublet.  The line ratios have been
corrected for dust extinction in the ORC4 central galaxy (with a global correction applied, not spaxel-by-spaxel, which is more uncertain); this correction only substantially
affects the \oiii \ to \oii \ ratio. 

The spatial variations seen for these line ratios is typically strongest near the edges
of the emission, where the errors on the flux increase, and may predominantly be due to noise.
However, the higher values in the \nii \ 6583 \AA / H$\alpha$ flux ratio along the northern
edge of the nebula may be real as the elevated line ratios can be seen by eye in the reduced
datacube in this region.  While the \oiii \ 5007 \AA / \oii \ 3727 \AA \ doublet flux ratio has
somewhat increased values in the center of the galaxy, the other line ratios do not display
substantially higher flux ratios in the galaxy center as might be expected from AGN
photoionization for many of these particular strong lines \citep{Groves04}. Overall, there
are not strong, coherent spatial gradients observed consistently across multiple line
ratios for this source.

Figure \ref{fig:bpt_sdss} presents four widely-used BPT-like diagrams for
spaxels in the ORC4 central galaxy (blue points with error bars) compared to
SDSS galaxies (contours).
The line ratios correspond to spaxels within a radius of 7 kpc from the
galaxy center. The observed line ratios for the ORC4 central galaxy are
presented without extinction correction, to maintain consistency with the
public SDSS line ratios, which are also uncorrected for extinction.

The upper left panel of Figure \ref{fig:bpt_sdss} shows the
\OIIIHb \ versus \NIIHa \ BPT diagram \citep{Baldwin81, Veilleux87}.
For comparison, SDSS sources are shown with contours, including 
all DR7 sources with $S/N>3$ in each of the four lines used to construct
the diagram. The dashed  green line indicates the local empirical 
division between star-forming galaxies and AGN from \citet{Kauffmann03}, while
the dot-dash green line indicates the local theoretical 
``maximum'' allowed starburst galaxy in \citet{Kewley01}. The dotted
 green line indicates the LINER demarcation line from \citet{Cid-Fernandes10}.
The ORC4 central galaxy lies in the composite/LINER region of this diagram.

The upper right panel shows the \OIIIHb \ versus \SIIHa \ diagram \citep{Veilleux87}, 
where the  dot-dash  green line again indicates the theoretical 
``maximum'' allowed starburst galaxy in \citet{Kewley01}, and the dotted
 green line indicates the LINER demarcation line from \citet{Kewley06}.
Here again the ORC4 central galaxy lies in the LINER region of this diagram.
In particular, for its \OIIIHb \ ratio ORC4 has an elevated \SIIHa \ ratio.
The lower left panel shows the \OIIIHb \ versus \OIHa \ diagram, where 
the dot-dash green line is from \citet{Kewley01}.
We find little overlap with SDSS sources, and ORC4 exhibits an elevated \OIHa \ ratio for its \OIIIHb \ value. The lower right panel shows the \OIIIOII \
versus \NIISII \ diagram, where ORC4 overlaps with more SDSS sources, displaying low \OIIIOII \ and somewhat low \NIISII \ values. Across all diagrams, the ORC4 spaxels span a range of $\sim$0.5 dex in both \OIIIHb \ and \OIIIOII \ ratios, with less variation in the remaining line ratios.

%%%%%%%%%%%%%%%%%%%%%%%%%%%%%%%%%%%%%%%%%%%%%%%%%

\begin{figure*}[ht!]
\centering
\includegraphics[width=6.5in]{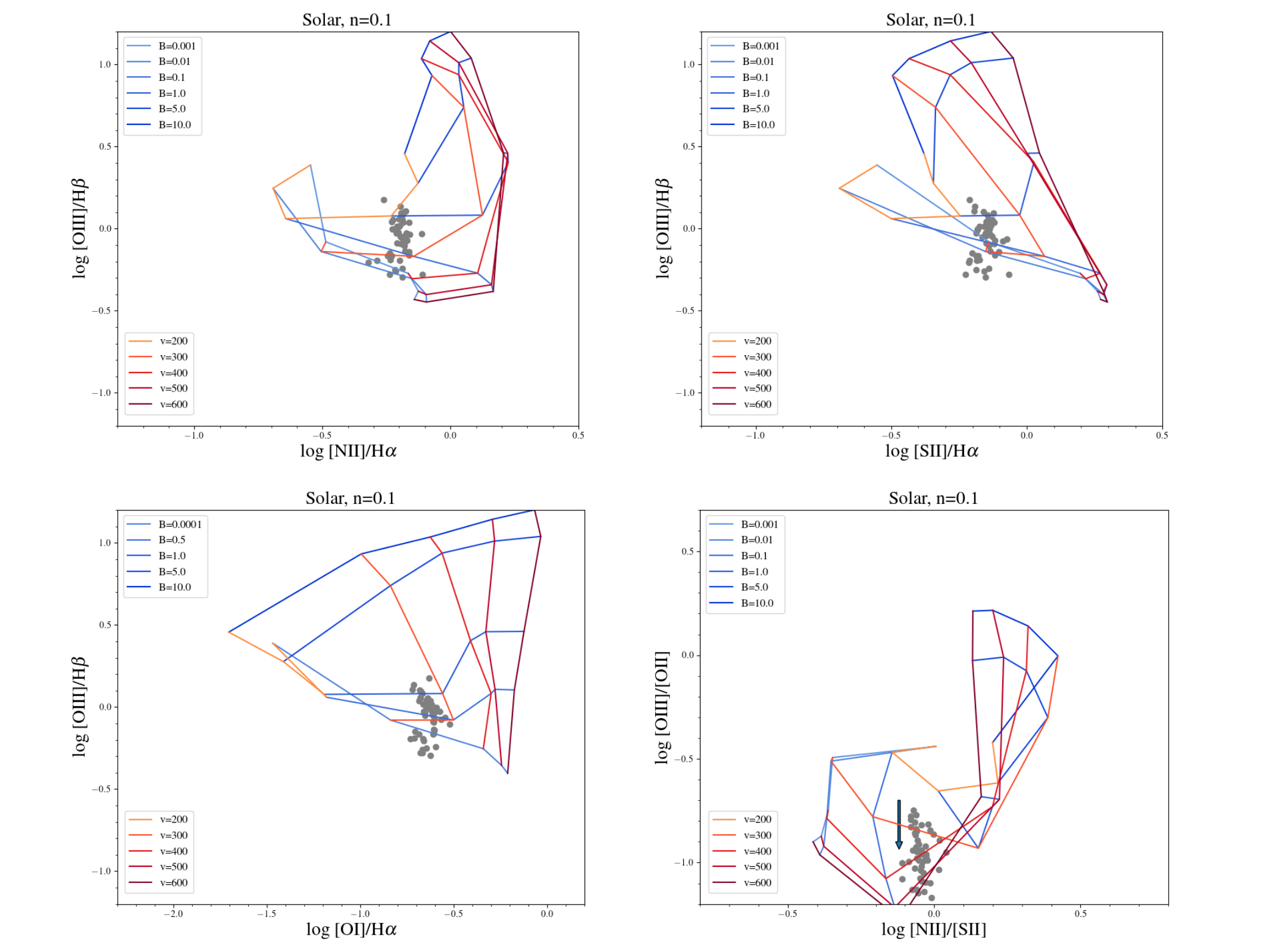}
\caption{A comparison of extinction-corrected line ratio diagrams for the ORC4 central galaxy with fast radiative shocks models of \cite{Allen08} for solar metallicity and a pre-shock gas density $n = 0.1 \ cm^{-3}$. The grids show varying values of velocity ($v$ in $km s^{-1}$) and
  magnetic field strength ($B$ in $\mu$G).
  As in Figure \ref{fig:agn_models2} the blue arrows indicate the magnitude and direction of the extinction correction applied in the ORC4 central galaxy.}
\label{fig:shock_models1}
\end{figure*}

%%%%%%%%%%%%%%%%%%%%%%%%%%%%%%%%%%%%%%%%%%%%%%%%%

\subsection{Comparison with AGN and Fast Shock Models} \label{sec:models}

We next compare the ORC4 line ratios with both AGN and shock photoionization
models, to clarify the nature of the ionizing radiation in the ORC4 central
galaxy. In this section we use extinction-corrected line ratios to compare to
models.  Applying this correction does not lead to a substantial shift in
any of the line ratios used here, except the \OIIIOII
\ line ratio which decreases by 0.20 dex.
The dark blue arrows in the lower right panels of the figures in this section
show the effect of extinction correction on the plotted line ratios, with the
arrow length showing the size of the correction.

Figure \ref{fig:agn_models2} compares the ORC4 results to the
\citet{Groves04} dusty, radiation pressure-dominated AGN model.  These
models assume a simple power-law ionizing spectrum, constant gas
pressure, and plane-parallel clouds, including both dust and the
effects of radiation pressure on it, which together stabilize the
ionization structure.  The free parameters are the AGN power-law index
($\alpha$) and ionization parameter (log $U$), which vary between $-2$
to $-1.2$ and $-4$ to 0, respectively.  In this model $U$ is a
dimensionless ionization parameter that is the ratio of the photon
density to the atomic density or the number of ionizing photons
relative to the hydrogen density.

We show model grids for an AGN with twice solar metallicity and an
electron density of $n_e = 100 \ cm^{-3}$, as this model provides the
best fit to the data of all of the AGN models. However, none of the
AGN models fit the ORC4 data well.  As can be seen in Figure
\ref{fig:agn_models2}, the ORC4 points lie fairly consistently near
the log $U = -3$ grids, however, the data are not consistent with any
values of $\alpha$ across all four line ratio spaces presented here.
While the \NIIHa \ and \SIIHa \ values are consistent with
$\alpha = -2.0$ at the given \OIIIHb \ values, and the \NIISII \
values are consistent with $\alpha$ ranging from $-2.0$ to $-1.4$ (as
the model grids are very close to each other in this space), the \OIHa
\ values are most consistent with $\alpha = -1.4$ and are a full
$\sim$0.8 dex higher at a given \OIIIHb \ ratio than the
$\alpha = -2.0$ model.  The other AGN models do not provide more
consistent fits to the data.

Notably, the \OIHa \ ratio is well-known to be sensitive to the
hardness of the ionizing spectrum, while low-ionization line ratios
such as \NIISII \ also depend on both the spectral shape and the
ionization parameter \citep[e.g.,][]{Blandford90, Kewley06,
  Koutsoumpou25}. Their disagreement with the AGN model grids strongly
suggests that AGN photoionization is not the dominant mechanism in the
ORC4 central galaxy.  The fact that some line ratios are consistent
with $\alpha = -2.0$ while others favor $\alpha = -1.4$ further
implies a composite ionization source or alternative ionizing
processes rather than a central AGN. Additionally, the \oi \ emission
line is more shock sensitive \citep{Rich11} and as such may indicate
the presence of shocks, which we consider next.

%%%%%%%%%%%%%%%%%%%%%%%%%%%%%%%%%%%%%%%%%%%%%%%%%

\begin{figure*}[ht!]
\centering
\includegraphics[width=6.5in]{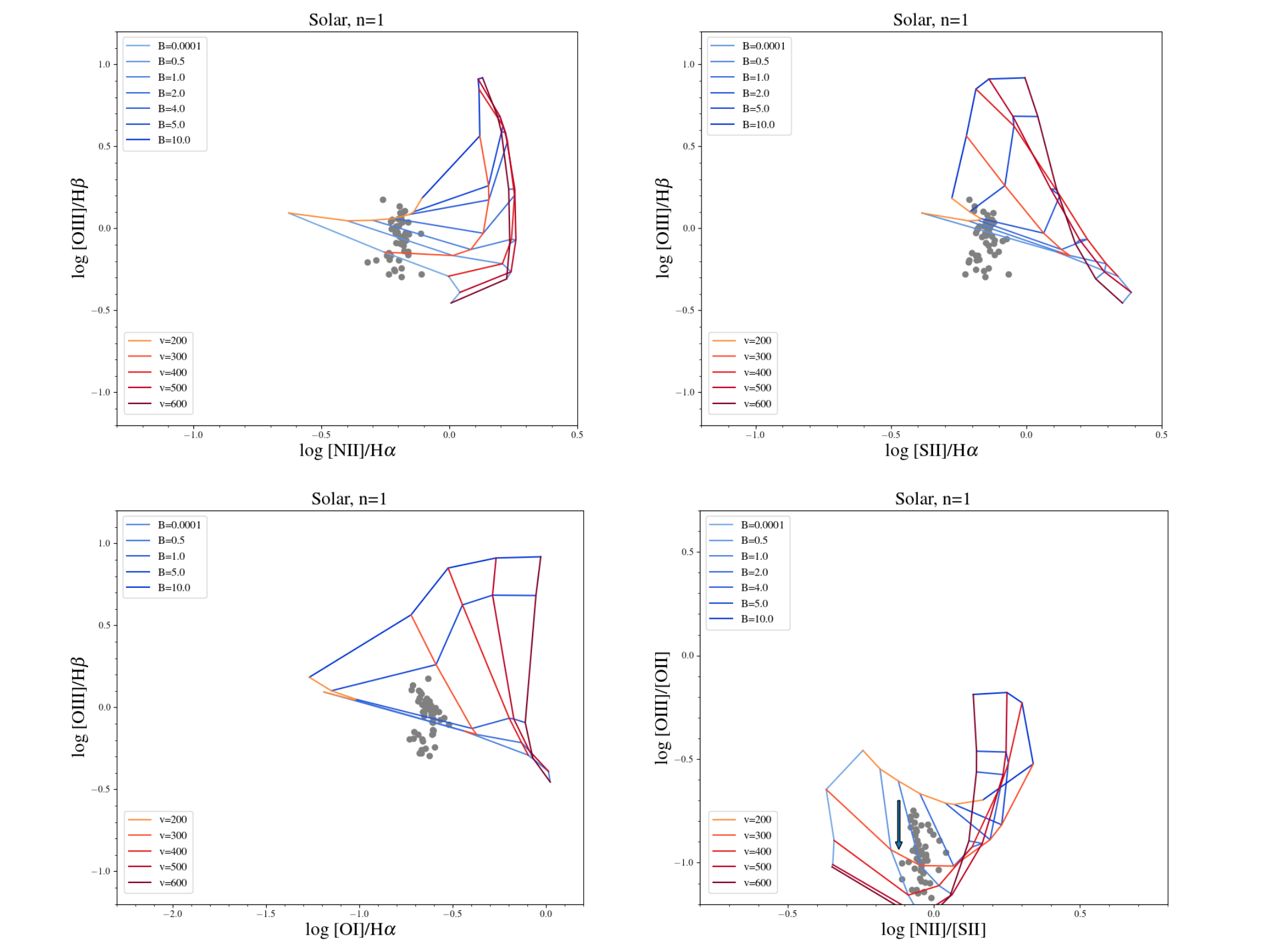}
\caption{A comparison of extinction-corrected line ratio diagrams for the ORC4 central galaxy with the \citet{Allen08} shocks models for solar metallicty and a pre-shock gas density $n = 1 \ cm^{-3}$.  As in Figure \ref{fig:shock_models1} the blue arrow in the lower right panel indicates the magnitude of the extinction correction applied.}
\label{fig:shock_models2}
\end{figure*}

\begin{figure*}[ht!]
\centering
\includegraphics[width=6.5in]{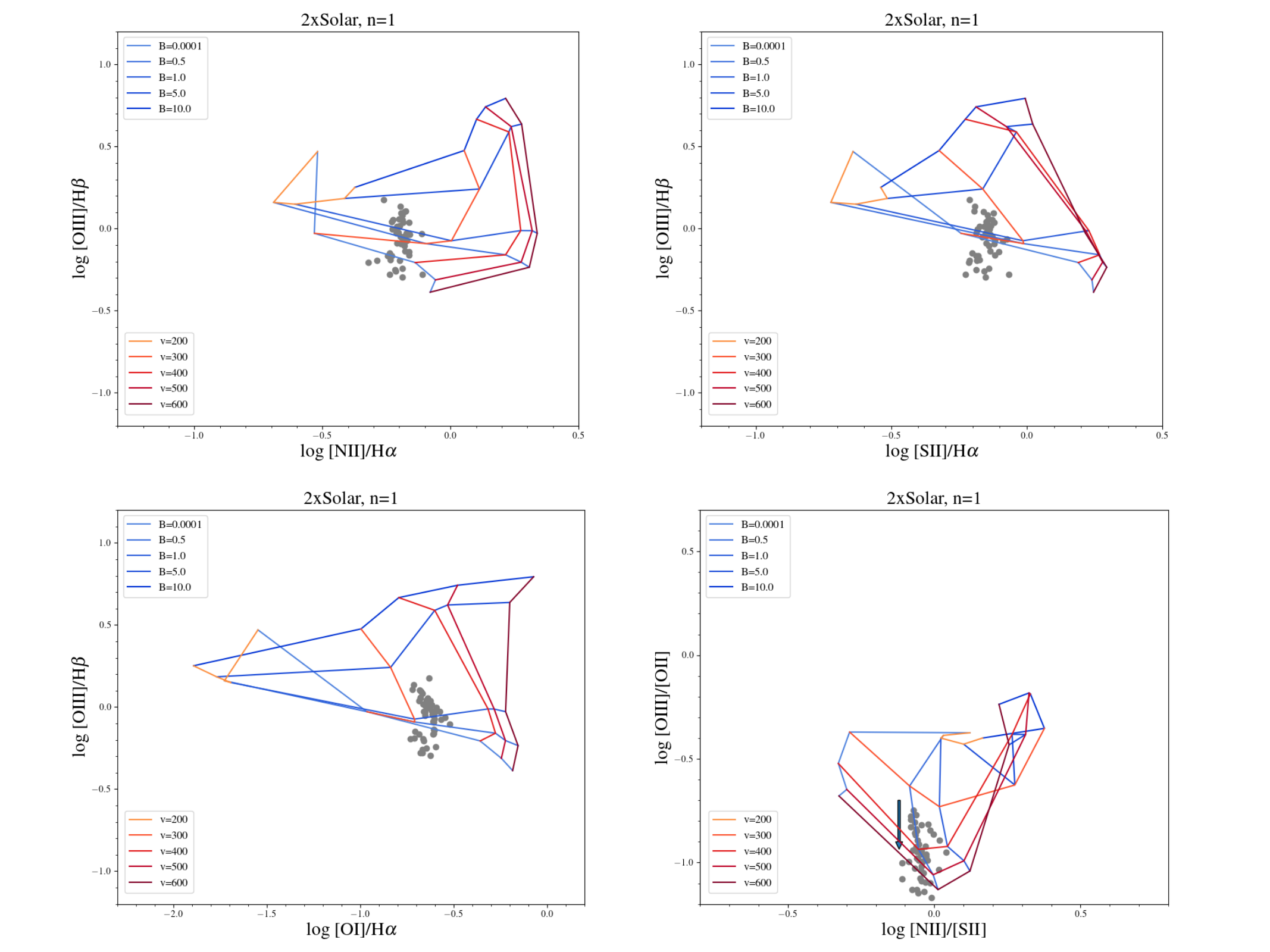}
\caption{A comparison of extinction-corrected line ratio diagrams for the ORC4 central galaxy with the \citet{Allen08} shocks models for twice solar metallicty and a pre-shock gas density $n = 1 \ cm^{-3}$.}
\label{fig:shock_models3}
\end{figure*}

%%%%%%%%%%%%%%%%%%%%%%%%%%%%%%%%%%%%%%%%%%%%%%%%% 

Figures \ref{fig:shock_models1}, \ref{fig:shock_models2}, and
\ref{fig:shock_models3} show comparisons of the ORC4 central galaxy line ratios
with the fast radiative shock models of
\citet{Allen08}. We first compare with shock photoionization only models (no precursor),
as these provide the best fit to the data. We compare with models of solar
metallicity and a pre-shock gas density of $n=0.1 \ cm^{-3}$ and $n=1 \ cm^{-3}$ and a
model of twice solar metallicity and $n=1 \ cm^{-3}$.
In general, the solar metallicity shock models fit the data well for
shock velocities of $v=200-300$ \kms \ and magnetic field strengths of
$B\sim0.1-1 \mu$G, while the twice metallicity shock models fit the
data at higher velocities $v=300-500$ \kms \ and similar magnetic
field strengths.  All four strong line ratio diagnostics examined here
provide fairly consistent fits to the shock models for these
velocities and magnetic field strengths.  In particular, the \OIHa \
versus \OIIIHb \ ratios observed in ORC4 are consistent with shock
model parameters that agree with the other line ratios, unlike the
case for the AGN models.  Of the three shock models presented, none
stands out as providing a substantially better fit to the data than
the others.  However, the observed velocity dispersion of gas is often
used a proxy for the shock model velocity \citep[e.g.,][]{Rupke23},
and if we interpret the observed $\sigma$ as the shock velocity, then
the solar models provide the best fit.  It is notable that in
  general the best fit shock velocities of $v=200-300$ \kms \ match
  well the observed velocity dispersion of the gas, $\sigma\sim200$
  \kms.  As noted above, the extinction correction primarily affects
the \OIIIOII \ ratio, and the exact value of the color excess is
somewhat uncertain. However, the uncorrected observed ratios are also
well matched by the same shock models such that applying the
extinction correction does not substantially change the results.

 We compare as well with the shock+precursor models of
  \citet{Allen08}. While the best fit shock velocities are on the
  lower end of the fast shocks, they have speeds at which there may be
  a non-negligible contribution from the photoionized precursor gas.
  The best fit shock+precursor model has twice solar metallicity and
  $n=1 \ cm^{-3}$ and has velocities of $v=300-350$ \kms \ and
  magnetic field strengths of $B\sim0.1-1 \mu$G. This model matches
  the observed \OIIIHb, \NIIHa, \SIIHa, and \NIISII \ ratios
  well. However, the predicted \OIIIOII \ ratio is higher than
  observed; this ratio is the most impacted by the dust correction.

It is notable for that these shock models in the velocity ranges
relevant here, the \OIIIHb \ and \OIIIOII \ ratios tend to decrease as
velocity increases. This could lead to an interpretation of the spread
in the ORC4 central galaxy line ratios observed here across different
spaxels as being due to a velocity difference across the galaxy.
However, the observed spread in \OIHa \ values is low, and this is the
lower ionization ratio that is most sensitive to shock velocity, such
that this argues against the observed spread in the spaxel-by-spaxel
variation being due to velocity differences. It is unclear what is
causing the vertical dispersion seen in these line ratio diagrams for
the ORC4 data points, though Figure \ref{fig:line_ratio_maps} shows
that it is not strongly correlated with radius or spatial location
within the nebula.

\section{Discussion} \label{sec:discuss}

The motivation for this work is to understand the origin of the
extended line emission observed in the ORC4 central galaxy. We
  have updated previous optical spectroscopy on ORC4 with a deep,
  spatially-resolved map of the full suite of strong emission line
  ratios in the rest-frame optical (Section \ref{sec:line-ratios}). This
  has allowed us to leverage well-understood ionization
  models (Section \ref{sec:models}). Here we discuss the implications
  for the origin of the ORC4 optical nebula in the context of these standard
  ionization models and by comparison to other systems in which
  similar line ratios are observed.

\subsection{Interstellar Gas or an AGN Outflow}

The ``null hypothesis'' is that the \oii\ emission is from the
  galaxy's interstellar medium. \citet{Coil24} conclude that this is
inconsistent with the properties of the \oii\ nebula---including its
unusual physical extent, extremely high EW, and high velocity
dispersion---and with the old stellar population in the host galaxy.
There is no indication of on-going star formation in the ORC4
  central galaxy, and \citet{Coil24} showed that the radial surface
  brightness profile of the \oii \ emission, which is similar to the
  radial profile of other emission lines shown here, does not follow
  that of the stellar light. This points to ionization from stars not
  being the dominant ionization source.

 A second possibility is excitation by an AGN. \citet{Coil24} and
\citet{Rupke24} conclude that, while a weak AGN may be present in the
ORC4 central galaxy, it is likely not the source of the strong line
emission, and \citet{Rupke24} estimate that if an AGN is present it
has a low Eddington ratio of log $L/L_{Edd} = $-3 to
-4. \citet{Coil24} also show that the \oii \ luminosity is extremely
high given the radio luminosity in this source, when compared to AGN.

Here we conclude as well that an AGN, if present, is not likely to be
the main source of photoionization. The AGN models cannot jointly
  fit the strong lines with consistent parameters, with the \OIHa \
  ratio in particular indicating a softer ionizing continuum than the
  other lines (Figure \ref{fig:agn_models2}). Any AGN would have to be
  low-luminosity to reproduce the optical line ratios and radio
  emission, but would then have difficulty exciting the observed
  emission lines, whose luminosities sum to almost 10$^{43}$ erg
  s$^{-1}$ (Table 1). Finally, the line ratio spatial maps (Figure
\ref{fig:line_ratio_maps}) do not show strongly elevated values in the
very center of the galaxy, nor are elevated line ratios observed in
the shape of a collimated bicone, as is common in galaxies with an AGN
ionization cone or outflow.

 The LINER-like line ratios observed in ellipticals---like those
  seen in the ORC4 central galaxy---can alternatively be interpreted
  as ionization of the interstellar medium (ISM) by older stellar
  populations. In this case, post-asymptotic giant branch stars are
  the best candidate \citep[e.g.,][]{Sarzi10}. However, the line
  emission in ORC4 is much brighter than in these galaxies and the radial profile of the ionized gas does not trace that of the stars 
  \citep{Coil24}.

\subsection{The Reverse Shock Model}

Overall, the fast radiative shock models of \citet{Allen08} provide a
good match to the extinction-corrected line ratios for solar and
twice-solar metallicity, with shock velocities of $\sim$300
\kms. Importantly, these models are consistent not only with global
observed line ratios but also with the spatially-resolved features in
ORC4.  The \oii \ emission detected to low surface brightness
  levels is spatially extended in the northwest, precisely where the
shock models predict enhanced emission due to turbulent or infalling
gas. Additionally, the high velocity dispersions ($\sigma\sim150-250$
\kms) measured across the nebula, and the asymmetric velocity
gradients seen in the \oii \ maps, are naturally explained by shocks
propagating through a disturbed or multi-phase medium. These spatial
correlations strengthen the argument that shocks, rather than AGN
photoionization, dominate the excitation of the ionized gas in the
central galaxy.

This scenario would appear to be consistent with the model presented
in \citet{Coil24} for ORC4 in which the same energetic event that
created the large-scale radio emission which identified the source as
an ORC also created the observed line emission in the central
galaxy. In this model a starburst-driven wind creates both a shock on
large scales from the forward-moving shock and also create a reverse
shock, often called a wind shock, on smaller scales.  Initially a
starburst episode drives an outflow to large scales and once the
starburst has ceased then the gas heated by the reverse shock falls
back towards the central galaxy \citep{Lochhaas18}. The shocked wind
fills the region in and around the galaxy that was initially cleared
by the outward-moving forward shock and can create a turbulent medium
full of additional shocks by interacting with gas in and around the
galaxy \citep{Dolag23}. This gas can then radiatively cool to produce
gas emission on the scale of the galaxy.  The forward shock can
continue towards larger scales and create the synchrotron emission
that results in the large-scale radio ring.

With the new KCWI+KCRM data presented here we find that this model
remains viable.  Specifically, the gas kinematics are consistent with
expections from infalling and/or turbulent gas, given the asymmetric
velocity gradients, predominantly redshifted velocity centroids, and
high velocity widths of the ionized and neutral gas.  The predominance
of the \oii \ emission is consistent with being enhanced by shocks
\citep{Allen08} and/or could be due to mixing between the hot and cool
gas phases at boundary layers due to shocks
\citep{Gronke20,Fielding22}. Additionally, a kinematic offset between
the \oii \ emission and other emission lines such as \oiii \ and
H$\alpha$ could arise if \oii \ is enhanced by shocks, such that the
different lines are tracing somewhat different gas phases.
\citet{Dolag23} present high-resolution cosmological zoom-in
simulations of a galaxy merger and find that internal shocks match
observations of ORCs including their physical extent, inferred Mach
numbers, and lack of star formation in the central galaxy. These
simulations show that multiple shock fronts result on different scales
within and around the central galaxy.

The new results presented here also match well the simulations of
\citet{Coil24}, in which a starburst-driven wind produced gas that,
after the wind shut off, returned to the galaxy at $\sim300$ \kms. The
galaxy itself was not modeled, but it is expected that this returning
gas would collide with itself and any remaining ISM to produce many
chaotic shocks and seed a highly turbulent region. The shock velocity
and turbulent velocity would be similar to the velocity of the
returning gas that drives the shocks and turbulence, so the measured
\oii\ $\sigma$ of $150-200$ \kms \ and inferred shock velocity of
$200-300$ \kms \ from the shock models are both consistent with the
\citet{Coil24} simulations. These simulations are spherically
symmetric and so cannot capture the velocity gradients across the
\oii\ nebula, but any asymmetries in a real system, either of the
starburst itself $\sim1$ Gyr ago or the ISM gas at the time of the
burst or the present time of observation, could drive instabilities or
affect the geometry of the returning gas to produce an asymmetric
velocity field.

 The shock model also constrains the energy required to power the
  nebula. \citet{Dopita95} provide a relationship between the
  line emission from shocks and the shock velocity and
  density. \citet{Rupke23} combine these to relate the total
  luminosity radiated by the shocks to the H$\alpha$ emission (their
  Equation 1). Using the unextincted line luminosity from Table 1, we
  calculate that the shocks radiate 10$^{44}$ erg s$^{-1}$ for a shock
  velocity of 200--300 km s$^{-1}$. For a returning wind falling from
  100 kpc to 5--10 kpc, this luminosity would require a mass inflow rate of
  $(1–2)\times10^3M_\odot$ yr$^{-1}$
  under the assumption that all the gravitational
  potential energy of the inflowing gas is converted into shock luminosity.

  We can use the simulations of \citet{Coil24} to calculate the kinetic
  energy of the returning wind more directly: the returning material has a
  total mass of $\sim2\times10^{11}M_\odot$ and spends $\sim200$ Myr
  impacting the galaxy at 250 km s$^{-1}$ (consistent with the $1000 M_\odot$
  yr$^{-1}$ expectation from converting gravitational potential energy directly to
  shock luminosity). This provides $10^{43}$ erg s$^{-1}$ of kinetic
  luminosity to the galaxy that is converted into shock luminosity when
  the gas shocks upon its return, somewhat smaller than the $10^{44}$ erg
  s$^{-1}$ shock luminosity. The discrepancy can likely be resolved with 
  different parameters for the initial wind bubble; we stress that the choices
  made for the \citet{Coil24} simulation are not the only parameters that can 
  produce both a large-scale ORC and returning, shocking gas that generates
  widespread and strong \oii\ emission, and different parameter choices 
  can affect the kinetic luminosity of the returning gas.

\subsection{Comparison to Other Systems}

There are other galaxies and systems with similar optical line
  ratios to those observed here. Comparing to these may shed further
  light on the origin of the emission in the ORC4 central galaxy.

  \citet{Rupke23} found that the optical line ratios observed in the
  large-scale (r$\sim$20-50 kpc) outflow in the Makani galaxy are also
  best matched by fast shocks from \citet{Allen08} for shock-only and
  shock plus precursor models with velocities of $v\sim200-250$ \kms \
  and low magnetic field strengths, similar to what we find for the
  ORC4 central galaxy. Similar to ORC4, they found that the \OIIIOII \
  ratios were lower than the model predictions. They conclude that
  this may be due to only partial pre-ionization of the precursor, as
  assumed in slow shock models \citep{Rich10, Dopita17} or additional
  post-shock physics. In Makani, the observed velocity dispersion of
  the gas also matched the model shock velocity of $\sim$200 \kms
  \citep{Rupke23}.

  The line ratios observed in the ORC4 central galaxy are also similar to
  those seen in brightest cluster galaxies (BCGs). The optical line
  ratios in cooling core clusters are generally consistent with
  LINER-like emission
  \citep[e.g.,][]{Voit97,Crawford99,Edwards07,McDonald12,Pagotto21}
  and have high \NIIHa, \SIIHa, and \OIHa \ ratios. While only
  $\sim$15\% of BCGs have optical line emission, this fraction rises
  to 71\% for BCGs in cooling flow clusters \citep{Edwards07}, which
  points to a connection to the hot cluster atmosphere. The morphology
  of this warm ionized gas is often highly asymmetric and extends in
  long, narrow filaments traced up to 50-100 kpc from the central
  galaxy \citep[e.g.,][]{Salome11,McDonald15}. These filaments
  frequently extend towards and around radio lobes and/or X-ray
  cavities resulting from previous AGN outflows
  \citep[e.g.,][]{Tremblay15}.

  For BCGs, it is not clear if there is a single dominant ionization
  mechanism, and the ionization mechanism for the gas in filaments may
  differ from that in the central galaxy. Some BCGs display signatures
  of recent or on-going star formation in the central galaxy
  \citep[e.g.,][]{Liu12,Gingras25} and/or AGN activity that likely
  contributes to the gas ionization \citep[e.g.,][]{Crawford99}. The
  observed \Ha \ line luminosities are much higher than would be
  expected due to direct cooling of the hot cluster gas
  \citep[e.g.,][]{McDonald12}, which indicates that the line emission
  is not primarily due to recombination from cooling, though
  reprocessing of energetic particles or EUV or X-rays could play a
  role \citep{Polles21}. Shocks are also thought to play a role, though
  they may not be the dominant or sole mechanism. \citet{Donahue00} found that
  strong shocks destroy H$_2$ and in BCGs the observed ratios of H$_2$
  to \Ha \ and X-rays are higher than predicted by strong shocks
  \citep[see also][]{Heckman89, Voit97}. Additionally, the lack of
  \oiii \ 4363 \AA \ emission in BCGs rules out fast shocks, though
  slow shocks may contribute to the gas ionization.

  While the ORC4 system has potential overlap with central galaxies of
  groups, it is unlikely to be a cluster BCG. As such, the variety of
  physical processes at play in clusters that impact the line
  ratios of BCGs may not be relevant for the ORC4 central galaxy. The
  stellar mass of the ORC4 central galaxy, \logmass $=11.3$,
  corresponds to a wide range of possible halo masses \citep[see
  Figure 3 of ][]{Wechsler18} from \logmhalo$\sim12-15$. However, the
  ORC4 central galaxy does not appear to reside at the center of an
  overdensity of galaxies \citep{Norris21b}. Morphologically, the
  ionized gas in ORC4 is not similar to that in BCGs; in ORC4 the gas
  is spherically symmetric and does not have a filamentary structure.

  The presence of a nearby galaxy with stellar mass \logmass$\sim10.6$
  \citep{Coil24} is consistent with the possibility of ORC4 being in a
  fossil group, though X-ray data are needed to confirm this \citep[][
  and references therein]{Zarattini14}. Fossil group brightest group
  galaxies typically have somewhat higher stellar mass than ORC4
  (\ensuremath{\log\mass/\msun}
$\sim$11.5) and roughly half contain optical emission lines
  \citep{Chu23}. Thus, we consider whether or not there could be a
  contribution to the ionization mechanism in ORC4 from cooling of hot
  gas in the halo. While the halo mass is unknown and X-ray data does
  not exist to infer a cooling rate, the H$\alpha$
  luminosity for ORC4 is similar to that of BCGs \citep{McDonald12},
  for which the line luminosity is too high to be due solely to
  cooling. Therefore cooling from a hot atmosphere is likely
  subdominant for ORC4, given that the cooling rate scales with
  group/cluster mass \citep{McDonald11}.

  Finally, the line ratios observed in ORC4 are similar to those seen
  in some parts of jellyfish galaxies \citep{Poggianti19}, which are cluster members
  undergoing extreme ram pressure stripping. In these systems, the
  correlation between the \Ha \ luminosity and X-ray luminosity point
  to a local ionization mechanism \citep{Olivares25}. The physical
  environment and morphology of the ORC4 nebula are not consistent
  with it being such a system. It is worth noting that some shocks are
  certainly present in these galaxies \citep[e.g.,][]{Wong14}.
  However, much of the clumpy emission in these galaxies is
  in fact more consistent with young star formation, unlike ORC4, and
  this clumpy emission dominates the warm ionized gas \citep{Poggianti19}.

  Ultimately, the similarity in line ratios among these systems and
  ORC4 points to similar gas physics on the small scales where the
  lines are emitted. All of these situations have dynamic,
  multiphase gas undergoing cooling and mixing, even though the origin
  of the cooling and mixing may arise from a different set of
  circumstances. This may be even stronger evidence for the returning
  wind model. In this scenario, once the returning wind hits the galaxy
  it likely becomes highly
  turbulent, generates shocks, and participates in cooling, all of
  which are small-scale processes that also occur in jellyfish galaxy
  tails and cool-core clusters. Thus, it is quite reasonable
  that the line ratios would be similar to these other systems,
  highlighting that line ratios alone are not sufficient to distinguish among
  origin scenarios.

\section{Summary and Conclusions}

In this paper we present new Keck/KCWI+KCRM integral field
spectroscopic data on the central galaxy in the odd radio circle ORC4.
We detect line emission from multiple strong optical lines including
\mgii, \oii, H$\beta$, \oiii, \oi, \nii, H$\alpha$, and \sii.

The morphology, extent, kinematics, luminosity, and EW of the \oii \
emission in ORC4 detected with integral field unit (IFU) data, as well
as the LINER-like line ratios observed for the redder optical lines
with GMOS long-slit data, are all consistent with a scenario in which
the ionized gas arises  predominantly from shocks near the galaxy associated with the
subsequent infall of material resulting from a larger-scale
forward-moving shock.
The line emission likely arises from a combination of gas cooling and mixing, resulting from a reverse shock.
Therefore both the optical and radio emission,
while observed on very different scales, may result from the same
dramatic event.  Massive starburst galaxies due to major mergers are
known to have very fast outflowing winds; these could be the
precursors to ORCs, as they drive winds with high mass outflow rates
($>200 M_\odot$/yr) to circumgalactic scales ($>$20 kpc), which are
needed to produce the observed synchrotron rings
\citep[e.g.,]{Rupke19, Perrotta23}.  \citet{Coil24} speculate that
ORCs may therefore reveal a later stage of these massive starburst
wind galaxies, after the wind has shut off and gas has fallen back
towards the galaxy.

The primary findings from this analysis are:

\begin{enumerate}

\item The line emission from \oii \ is both higher luminosity and
   detected to larger radius than that observed for the other strong optical emission lines. We confirm that the \oii \ restframe EW is very high across the central galaxy ($\sim$60 \AA) and is particularly enhanced ($>$100 \AA) 10-15 kpc to the west of the nucleus. 

\item We detect strong \nii, \oi, and \sii \ emission, while the \oiii \ emission
  is relatively weak in the ORC4 central galaxy.

\item The gas kinematics show strong spatial asymmetries across the central galaxy with high velocity gradients of $>100$ \kms \ and a high velocity dispersion of $\sim200$ \kms.
  
\item The high luminosities and EWs observed in the ORC4 central galaxy, as well as the optical line ratios, are not typical of massive early-type galaxies and likely do not
  correspond to emission from either young or old stars or photoionization from an AGN. The line ratios are not substantially spatially elevated near the nucleus of the galaxy. 

\item The observed line ratios are most consistent with shock photoionization from shocks with velocities of $\sim200-300$ \kms.  These results are consistent with spatially-extended shocks existing throughout the ORC4 central galaxy.

\end{enumerate}

It is perhaps not surprising that spatially-extended shocks would
exist in the central galaxy of an ORC, if the large-scale radio
emission is the result of an outward-moving shock or shell, as appears
to be the case for some ORCs \citep{Norris22}.  The star formation
history of the ORC4 central galaxy indicates a starburst $\sim$1 Gyr
ago, which may have been due to a merger and subsequent starburst
episode.  As extreme starbursts are known to drive outflowing galactic
winds \citep{Rupke19, Perrotta24}, it is plausible that the
large-scale radio emission and the shocked gas in the ORC4 central
galaxy are the result of a past major merger.

While ORCs may be a rare and possibly transient phenomenon, they have
great potential as a probe of the physical conditions in the
circumgalactic medium of their host galaxies.  Future data across the
electromagnetic spectrum would be highly valuable, from X-ray emission
probing the hot gas in and around the central galaxies to integral
field optical data on the neutral and ionized gas to observations of
H$_2$ to probe shock models.  Additional information on the dust
properties of the gas would also be useful to help understand not only
the current properties of ORCs but the past events that created the
large-scale radio emission.  Additionally, high-resolution
hydrodynamical simulations that can resolve shocks and include the
properties of the host galaxies and their strong starburst feedback
events would strongly elucidate the nature of these objects.

\section*{Acknowledgments}
ALC is supported by the Ingrid and Joseph W. Hibben endowed chair at UC San Diego. Support for CL was provided by NASA through the NASA Hubble Fellowship grant HST-HF2-51538.001-A awarded by the Space Telescope Science Institute, which is operated by the Association of Universities for Research in Astronomy, Inc., for NASA, under contract NAS5-26555.
Some of the data presented herein were obtained at Keck Observatory, which is a
private 501(c)3 non-profit organization operated as a scientific partnership
among the California Institute of Technology, the University of California, and
the National Aeronautics and Space Administration. The Observatory was made possible
by the generous financial support of the W. M. Keck Foundation. The authors wish
to recognize and acknowledge the very significant cultural role and reverence that
the summit of Maunakea has always had within the Native Hawaiian community. We are
most fortunate to have the opportunity to conduct observations from this mountain.

\vspace{5mm}
\facilities{Keck:II (KCWI)}

\vspace{5mm}
\software{\texttt{astropy} \citep{Astropy13, Astropy18, Astropy22}, \texttt{matplotlib} \citep{matplotlib}, \texttt{NumPy} \citep{numpy}, texttt{IFSFIT} \citep{Rupke14, Rupke17}, \texttt{ppxf} \citep{Cappellari12, Cappellari17}}

\bibliography{references}{}
\bibliographystyle{aasjournal}

\end{document}